\documentclass[acmsmall,manuscript,screen]{acmart}

\AtBeginDocument{%
	\providecommand\BibTeX{{%
			\normalfont B\kern-0.5em{\scshape i\kern-0.25em b}\kern-0.8em\TeX}}}

\usepackage{booktabs} %
\usepackage[flushleft]{threeparttable}
\usepackage[utf8]{inputenc}
\usepackage{float}
\usepackage{gensymb}
\usepackage{mathrsfs}
\usepackage{amsfonts}
\usepackage{amsmath}
\usepackage{ifthen}
\usepackage{multirow}
\usepackage{caption}
\usepackage{listings}
\usepackage{hyperref}
\usepackage{lipsum} 
\usepackage{url,moreverb,xspace}
\usepackage{enumitem}
\usepackage{array,graphicx}
\usepackage{soul}
\usepackage{balance}
\usepackage{pifont}
\usepackage{etoolbox,siunitx}
\usepackage{nicefrac}
\usepackage{mathtools}

\usepackage{tikz}
\usepackage{pgfplots}
\usepackage[titletoc]{appendix}
\pgfplotsset{compat=1.16}
\usepgfplotslibrary{statistics}
\usetikzlibrary{fadings}
\usetikzlibrary{arrows.meta,shapes.geometric,shadows}

\definecolor{ao}{rgb}{0.0, 0.5, 0.0}

\usepackage[most]{tcolorbox}

\usepackage{xcolor,pifont}
\usepackage{color, colortbl}
\newcommand*\colourcheck[1]{%
	\expandafter\newcommand\csname #1check\endcsname{\textcolor{#1}{\ding{52}}}%
}
\newcommand*\colourcross[1]{%
	\expandafter\newcommand\csname #1cross\endcsname{\textcolor{#1}{\ding{56}}}%
}

\usepackage[lined,boxruled,norelsize,linesnumbered]{algorithm2e}

\makeatletter
\renewcommand{\ForEach}[3]{\ForEach{#1}{#2}{#3}}
\SetKwFor{ForEach}{for each}{do}{end for}
\makeatother

\def\HiLi{\leavevmode\rlap{\hbox to \hsize{\color{gray!35}\leaders\hrule height .8\baselineskip depth .5ex\hfill}}}

\definecolor{lightgray}{gray}{0.85}
\definecolor{bananayellow}{rgb}{1.0, 0.88, 0.21}

\newcommand*\rot{\rotatebox{90}}

\usepackage{graphicx}
\usepackage{subcaption}
\usepackage{seqsplit}
\usepackage[normalem]{ulem}
\usepackage[final]{listofsymbols}

\usepackage{soul}
\usepackage{tikz}

\useunder{\uline}{\ul}{}

\newcommand{\tool}{\textsc{Indago}\xspace} %

\SetKwComment{Comment}{$\triangleright$\ }{}
\SetCommentSty{itshape}
\newboolean{showcomments}
\setboolean{showcomments}{true}
\ifthenelse{\boolean{showcomments}}
{\newcommand{\nb}[2] {
		\fcolorbox{black}{gray!20}{\bfseries\sffamily\scriptsize#1:}
		{\sf\small$\blacktriangleright$\textit{#2}$\blacktriangleleft$}
	}
}
{\newcommand{\nb}[2]{}
}

\definecolor{myyellow}{RGB}{255, 228, 26}
\newcommand{\COMMENT}[1]{}

\newcommand{\head}[1]{\noindent\textbf{#1.}}
\newcommand{\headcolumn}[1]{\noindent\textbf{#1:}}

\newcommand{\curl}[1]{\footnote{\url{#1}}}

\opensymdef
	\newsym[Timestep]{symt}{t}
	\newsym[State]{syms}{s}
	\newsym[Action]{syma}{a}
	\newsym[Reward]{symr}{r}
	\newsym[Policy]{sympi}{\pi}
	\newsym[Optimal policy]{sympis}{\pi^*}
	\newsym[Set of states]{symS}{\mathcal{S}}
	\newsym[Set of actions]{symA}{\mathcal{A}}
	\newsym[Value function]{symv}{v_{\pi}}
	\newsym[Action value function]{symq}{q_{\pi}}
	\newsym[Optimal action value function]{symqs}{q^*_{\pi}}
	\newsym[Environment configuration]{syme}{e}
	\newsym[Environment configuration where the agent fails]{symef}{e_f}
	\newsym[Set of environment configuration where the agent fails]{symEf}{E_f}
	\newsym[Class label when training the surrogate model]{symc}{c}
	\newsym[Number of classes when training the surrogate model]{symCbar}{|C|}
	\newsym[Interaction dataset]{symD}{D}
	\newsym[Weight vector]{symW}{W}
	\newsym[Set of environment configurations]{symE}{E}
	\newsym[Set of real numbers]{symR}{\mathbb{R}}
	\newsym[Surrogate model]{symf}{f}
	\newsym[Optimal surrogate model]{symfs}{f^*}
	\newsym[Neighborhood size in hill climbing algorithm]{symNS}{NS}
	\newsym[Failure prediction]{symfp}{fp}
	\newsym[Population size in genetic algorithm]{symPS}{PS}
	\newsym[Crossover rate in genetic algorithm]{symcr}{cr}
	\newsym[Offspring environment configuration in genetic algorithm]{symoe}{oe}
	\newsym[Parent environment configuration in genetic algorithm]{sympe}{pe}
	\newsym[Environment configuration after crossover]{symce}{ce}
	\newsym[Small constant value used in the initialization phase of the Humanoid environment]{symm}{m}
	\newsym[Number of environment configurations for the training of the surrogate model]{symN}{N}
	\newsym[Set of class targets for the training of the surrogate model]{symY}{Y}
	\newsym[Number of environment configurations generated by each failure search approach]{symT}{T}
	\newsym[Search budget in seconds, for each failure search approach]{symB}{B}
	\newsym[Vargha-Delaney effect size]{symVD}{\hat{A}_{12}}
	\newsym[Significance level for the Mann Whitney U test]{symalpha}{\alpha}
	\newsym[Number of clusters for the k-means clustering algorithm]{symk}{k}
	\newsym[Optimal number of clusters for the k-means clustering algorithm]{symks}{k^*}
	\newsym[Generic failure search approach]{symAg}{A}
	\newsym[Cluster coverage metric]{symC}{C}
	\newsym[Function that measures if a certain cluster is covered by a given failure search approach]{symgamma}{\gamma}
	\newsym[Function that measures the probability of finding a failure of a certain failure search approach in a cluster]{symeta}{\eta}
	\newsym[Function that measures the number of failures generated by a given failure search approach in a cluster]{symF}{F}
	\newsym[Entropy metric]{symH}{H}
	\newsym[Failure class label]{symfc}{c_f}
	\newsym[Gini purity coefficient]{symG}{G}
	
\closesymdef

\makeatletter
\newcommand{\thickhline}{%
	\noalign {\ifnum 0=`}\fi \hrule height 1pt
	\futurelet \reserved@a \@xhline
}
\makeatother

\acmBooktitle{TOSEM}

\begin{document}

	\title{Testing of Deep Reinforcement Learning Agents with Surrogate Models}
	
	\author{Matteo Biagiola}
	\email{matteo.biagiola@usi.ch}
	\author{Paolo Tonella}
	\email{paolo.tonella@usi.ch}
	\affiliation{%
		\institution{Universit\`a della Svizzera italiana}
		\city{Lugano}
		\country{Switzerland}
	}

	\begin{abstract}

Deep Reinforcement Learning (DRL) has received a lot of attention from the research community in recent years. As the technology moves away from game playing to practical contexts, such as autonomous vehicles and robotics, it is crucial to evaluate the quality of DRL agents.

In this paper, we propose a search-based approach to test such agents. Our approach, implemented in a tool called \tool, trains a classifier on failure and non-failure environment (i.e., pass) configurations resulting from the DRL training process. The classifier is used at testing time as a surrogate model for the DRL agent execution in the environment, predicting the extent to which a given environment configuration induces a failure of the DRL agent under test. The failure prediction acts as a fitness function, guiding the generation towards failure environment configurations, while saving computation time by deferring the execution of the DRL agent in the environment to those configurations that are more likely to expose failures.

Experimental results show that our search-based approach finds 50\% more failures of the DRL agent than state-of-the-art techniques. Moreover, such failures are, on average, 78\% more diverse; similarly, the behaviors of the DRL agent induced by failure configurations are 74\% more diverse.

	\end{abstract}
	

	
	\setcopyright{acmlicensed}
	\acmJournal{TOSEM}
	\acmYear{2023} \acmVolume{1} \acmNumber{1} \acmArticle{1} \acmMonth{1} \acmPrice{15.00}\acmDOI{10.1145/3631970}

	\maketitle
	
	\section{Introduction} \label{sec:introduction}

Reinforcement Learning (RL) is a learning paradigm in which an agent interacts with the environment to complete a given task. Learning is driven by a reward signal returned to the agent by the environment. The ultimate goal of an agent is to learn a policy, i.e., a way to act in each environment state, that maximizes the amount of reward the agent earns in its lifetime.
The first RL algorithms were tabular~\cite{rlbible} and could handle tasks with finite and small environment states. With the advent of Deep Learning (DL), new algorithms (called Deep RL, or DRL for short) were proposed, which could deal with complex environment states (e.g., images)~\cite{dqn}. %

DRL has recently been applied in many practical contexts. An example is personalization, i.e., the problem of customizing a service to the needs of a particular user. For instance, Netflix uses DRL to choose which movie artwork to show to a user in order to maximize engagement~\cite{netflix-artwork-personalization}. Similarly, Microsoft developed Personalizer~\cite{microsoft-personalizer}, a service developers can use for content recommendation and ad placement. Meta proposed Horizon~\cite{facebook-horizon, facebook-reagent} (also called ReAgent), an open source applied DRL platform, employed to deliver personalized notifications to their users, replacing the previous system based on supervised learning.

Another practical context in which the DRL paradigm is applied is continuous control. For instance, the automaker Audi~\cite{audi-parking-rl} showcased that their 1:8 scale car could, using DRL, search for a parking place in an area of 9 square meters and park autonomously. Indeed, advancements in state representation learning and smooth exploration~\cite{srl-survey, raffin-smooth-real-robot, finnish-donkey} made it possible to train DRL agents directly on real robots. 
Moreover, in recent years,  simulators have become more realistic for a variety of tasks, besides achieving high parallelization thanks to GPU acceleration~\cite{isaac-gym, raisim}. In addition, domain randomization and learned actuator dynamics are reducing the sim-to-real gap in robotics research~\cite{legged-robots-learning-agile, legged-robots-learning-locomotion, learning-walk-minutes-parallel}.

Despite the growing prevalence of DRL agents in the real world, methodologies for testing such agents are still largely unexplored. On the other hand, DRL agents present some peculiar characteristics. Indeed, DRL agents are trained online since they interact with the environment to learn the optimal actions to perform the task. Specifically, in order to increase generalization, DRL agents are usually trained on randomized environment configurations~\cite{uesato-adversarial,everett-thesis} to prevent agents from memorizing how to behave in a particular instance of the environment (e.g., a self-driving car that drives only on a specific track).
Therefore, during training the DRL agent is presented with different environment configurations (e.g., different tracks) and it \textit{fails} in some while it \textit{succeeds} in others. 

The current state-of-the-practice to test DRL agents is to run them on a set of environment configurations generated at random~\cite{uesato-adversarial,gym-leaderboard}. However, testing a DRL agent on randomly generated environment configurations has two shortcomings. First, random generation~\cite{gym-leaderboard} is unlikely to expose failures. As a consequence, their absence might lead the developer to overestimate the capabilities of the DRL agent and to the deployment of an unsafe agent. Secondly, even finding \textit{challenging} environment configurations by random exploration is difficult and computationally expensive, since many executions are needed and each execution requires running the DRL agent in a simulator or in the real world.

On the other hand, the \textit{interactions} of the DRL agent with the environment during training provide clues about the weaknesses of the DRL agent that results from the training process. The intuition is that training failures are representative of \textit{critical} environment configurations even for the DRL agent once it has been trained, and could be used as guidance for the generation of new environment configurations that will likely challenge it. 

Our approach, implemented in a tool called \tool, considers the interaction data produced during the DRL training process as a labeled dataset to train a surrogate model --- i.e., a classifier --- on failure and non-failure (i.e., pass) environment configurations. 
Then, \tool uses such surrogate model as a proxy for the execution of the DRL agent in an environment with a newly generated configuration. \tool uses a search-based approach to maximize the failure prediction for an environment configuration given by the surrogate model, instead of random search~\cite{uesato-adversarial}. Indeed, random search needs to generate a large set of environment configurations in order to be effective, which makes it computationally expensive.
On the other hand, \tool uses a mutation operator guided by saliency-based input attribution~\cite{saliency-paper} to identify mutations that have the maximum influence on the failure prediction. This way, \tool turns a non-promising environment configuration into a critical one for the DRL agent under test, making the search more efficient. \tool executes the DRL agent under test only on the most promising environment configurations, i.e., those with the highest failure predictions, hence saving computation time while maximizing failure exposures.

Our paper makes the following contributions:

\headcolumn{1) Failure Search with Surrogate Models} in this paper we propose an approach that makes use of a surrogate model of the environment to guide failure search while automatically generating environment configurations for a DRL agent. In particular, we train a classifier on the training interaction data and use its output as a fitness function to maximize the failure prediction of a given environment configuration. Moreover, we use the saliency method to efficiently identify the most critical mutations for the given environment configurations.

\headcolumn{2) The \tool Tool} a practical tool, that implements the aforementioned approach, which we make publicly available~\cite{replication-package}. We also release three DRL agents trained on as many complex environments, as well as the required infrastructure to test them.

\headcolumn{3) Experimental Evaluation} we systematically compare different configurations of \tool with the state-of-the-art sampling approach~\cite{uesato-adversarial} that maximizes failure prediction by generating a large set of environment configurations. On three complex case studies, i.e., a parking task~\cite{highwayenv}, a walking humanoid~\cite{mujoco} and a self-driving car~\cite{sdsandbox}, our experiments show that, overall, \tool is able to find 50\% more failure environment configurations than sampling. Moreover, we introduce a clustering-based technique to measure the diversity of failure environment configurations triggered by the competing approaches. Experimental results show that the failure environment configurations found by \tool are 77\% more diverse than those generated by sampling. Moreover, the behaviours of the DRL agent induced by such environment configurations are 74\% more diverse.

	\section{Related Work} \label{sec:related-work}

Testing DRL agents is a rather unexplored area of research. Works that generate adversarial attacks~\cite{adversarial} have been proposed~\cite{adversarial-rl-generic-huang, adverarial-rl-tactics-lin}, showing that such DRL agents can be vulnerable to adversarial attacks, similarly to DL agents (i.e., agents trained using supervised learning). However, our approach is substantially different since it is not focused on perturbing the raw inputs of the DRL agent sensors but rather on generating configurations for the whole environment the DRL agent runs into. The most similar work to ours is that by Uesato et al.~\cite{uesato-adversarial}, who proposed the sampling approach we used as baseline in our experiments. Our results show that our search-based approach outperforms it both in terms of number of failures triggered and their diversity in all case studies.

Biagiola et al.~\cite{biagiola-plasticity} proposed an approach to test the adaptation capabilities of DRL agents. In particular, the training of a DRL agent is resumed with environment configurations that are different from those experienced by the DRL agent during the previous training phase. Then, the proposed approach builds the adaptation frontier of the DRL agent, separating the configurations in which the DRL agent under test is able to adapt from those where adaptation is unsuccessful. Our approach is similar, in the sense that we also generate environment configurations. However, we are interested in testing the DRL agent to find its weaknesses at testing time rather than its adaptation capabilities (i.e., we do not resume training).

Also related to our work is that by Ruderman et al.~\cite{everett-thesis}, who studied how to train and test agents in procedurally generated environments. In particular, they trained DRL agents on a set of procedurally generated mazes for a 3D navigation task. Then, at testing time, they adopt a local search process to modify generated mazes guided by the performance of the DRL agent. This process generates out-of-distribution mazes, i.e., environment configurations that are not possible under the training distribution. In our work, we minimize the computational cost of the search by using a surrogate model of the environment (i.e., a failure predictor), rather than executing the DRL agent under test in the environment to measure its performance. Indeed, executing the DRL agent in the environment at each search iteration becomes prohibitively expensive in environments more complex than mazes. Moreover, our approach does not generate out-of-distribution environment configurations since all environment configurations generated at testing time are subject to the same validity constraints of the environment configurations generated during the DRL training process. 

More recently, Tappler et al.~\cite{tappler-search} proposed a search approach to assess the quality of DRL agents. Their approach consists of searching for a reference trace that solves the RL task by sampling the environment. Such trace is built using a depth-first search algorithm and it is composed of all the states not part of the backtracking branches of the search. In particular, the search backtracks when it encounters an unsafe state, which is a state where the environment terminates the episode unsuccessfully. 
The states in the search graph preceding both an unsafe state and at least a successor non-terminal state, are called boundary states. The prefixes of the reference trace that end up in a boundary state are safety tests, since they are sequences of actions designed to bring the DRL agent under test into safety-critical situations. 
Our approach is complementary, since our goal is to generate new environment configurations to test the DRL agent, rather than evaluating it in the same environment configuration.

The literature in testing DL agents is quite rich and includes a multitude of works summarized in different surveys on the topic~\cite{2020-Humbatova-ICSE,zhang2020machine,foutse-survey}. Particularly relevant to our work are those that use search-based methods to generate test inputs~\cite{ase16-abdessalem,ase18-abdessalem,icse18-abdessalem,asfault,2020-Riccio-FSE}. However, our work is different since it specifically targets DRL agents by exploiting the interaction data a DRL agent produces during training, which is not possible in DL testing since DL agents are trained offline.

	\section{Background and Motivation} \label{sec:background}

\subsection{Reinforcement Learning}
\head{Fundamentals and Notations} Reinforcement Learning (RL) aims at learning a policy, which is a mapping from states to actions, in order to optimize a numerical reward signal~\cite{rlbible}. The agent that acts in the environment needs to discover what actions result in a high reward thorough trial and error without the presence of a supervisor. The main assumption of this learning paradigm is the so-called reward hypothesis~\cite{rlbible}, stating that training goals can be expressed as the maximization of the cumulative reward.

More formally, at each timestep \symt an RL agent receives a state \syms as input from the environment, and it has to decide the action \syma to take. The executed action triggers a change of state and results in a reward value \symr given to the agent by the environment. Assuming that the task we formalize as an RL problem is episodic, i.e., it terminates once certain conditions hold, the goal of an RL agent is to maximize the cumulative reward (i.e., often called return) of the episode.
In the general case, however, the RL objective is expressed as the maximization of the expected return since both the environment in which the agent operates and its policy can be stochastic. The main reason for the stochastic nature of a policy is due to a fundamental dilemma in RL, which is the exploration-exploitation dilemma. In fact, on the one hand, the agent has to exploit the actions already known to be rewarding but, on the other hand, it has to explore unknown actions that might result in even more reward. %

The most important component of an RL agent is its policy $\sympi: \symS \rightarrow \symA$, where \symA is a set of actions and \symS is a set of states.
The optimal policy \sympis tells the RL agent how to act in each state in order to maximize the expected return.
The optimal policy can  be learned directly or can be extracted from value functions, namely the state-value function $\symv(\syms)$ and the action-value function $\symq(\syms, \syma)$. The former quantifies the value of the state \syms, i.e., the expected return in \syms, whereas the latter quantifies the value of the state-action pair $(\syms, \syma)$. 
The difference between the two functions is that \symv provides the value of a state \syms by considering each possible action the agent can take in \syms (in other words, the average of the expected return in \syms for all the actions), while \symq considers the value of a state \syms for a particular action \syma. 
Both functions satisfy the Bellman equations~\cite{rlbible} which are recursive consistency equations relating the values of a state (or state-action pair) to the values of all the possible successor states (or state-action pairs). In particular, by solving the Bellman equation for \symq, we obtain \symqs from which we can extract \sympis by choosing in each state \syms the action \syma that maximizes \symqs.

\head{Deep RL Algorithms} Before deep learning, the RL problem was addressed using dynamic programming and approximate tabular methods, such as monte carlo methods and temporal difference learning~\cite{rlbible}. 
However, such methods are not applicable to problems where the state dimensionality is high (e.g., images) and/or the action space is continuous (e.g., the throttle in a self-driving car). The advent of deep learning made it possible to create Deep RL (DRL) algorithms that work on such complex practical scenarios~\cite{dqn}. 
In particular Deep RL (DRL) algorithms use non-linear function approximators, such as neural networks, to approximate high dimensional state and action spaces besides modeling the dynamics of the environment. Therefore, instead of having exact representations of policies \syms and value functions (\symv and \symq), neural networks can be used to approximate such quantities as well as to model the environment in which the agent operates.

In our experiments we consider model-free DRL algorithms which do not use a model of the environment. There exist different categories of model-free algorithms, based on how the RL problem is addressed. 
Specifically, policy gradients algorithms, of which PPO is a notable example~\cite{ppo}, directly solve the RL objective by representing the policy explicitly (i.e., with a neural network). Value-based algorithms, on the other hand, extract the policy by solving the Bellman equations, hence representing value functions explicitly. Examples of algorithms in such category are DQN and its improvements~\cite{dqn, double-q-learning, dueling-dqn, prioritized-dqn, rainbow-dqn, qrdqn}. Hybrid methods represent both policy and value functions to incorporate, the benefits of both policy gradients and value-based methods. SAC~\cite{sac} and TQC~\cite{tqc} are the state-of-the-art algorithms in this category. Belonging to a special category is the algorithm HER~\cite{her}, which was proposed as a wrapper on top of traditional DRL algorithms, to speed up learning in the context of goal-based tasks, e.g., parking a car from a starting position to a target parking spot.

\begin{figure}[t]
\centering

\includegraphics[trim=0cm 19cm 30cm 0cm, clip=true, scale=0.235]
{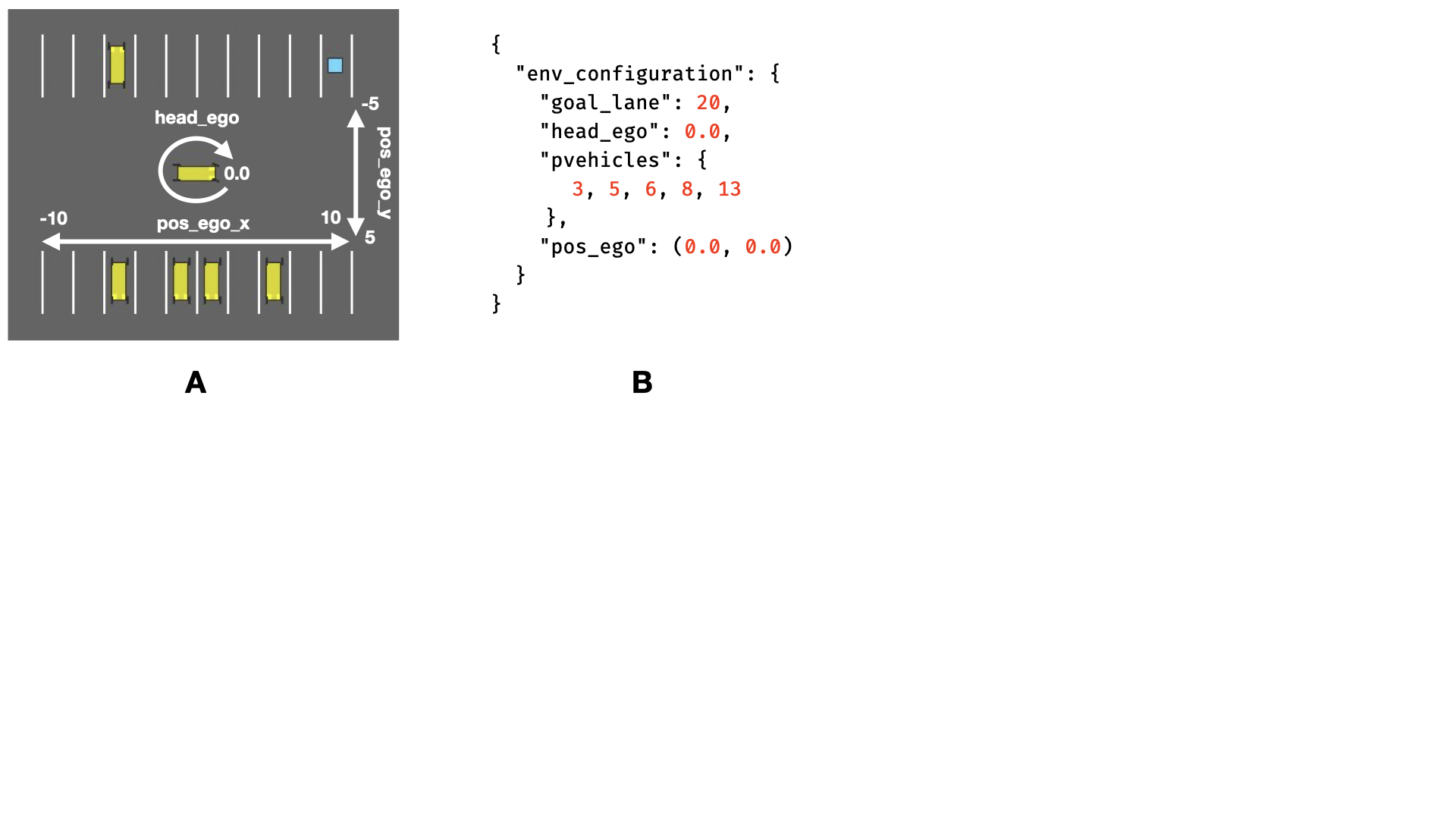}

\caption{An initial configuration of the \textit{Parking} environment in the \textit{HighwayEnv} simulator~\cite{highwayenv}. The left-hand side (\textbf{A}) shows how the configuration on the right-hand side (\textbf{B}) is rendered in the environment.} 
\label{fig:background:parking-environment} 
\end{figure}

\subsection{Motivating Example}

\autoref{fig:background:parking-environment} shows the \textit{Parking} environment, created by Leurent et al.~\cite{highwayenv}, in a particular configuration where the DRL agent needs to control the ego vehicle positioned at the center of the parking place. The left-hand side of the figure (i.e., \autoref{fig:background:parking-environment}.\textbf{A}) shows how the environment configuration on the right-hand side (i.e., \autoref{fig:background:parking-environment}.\textbf{B}) is rendered in the \textit{HighwayEnv} simulator.
In such environment configuration the ego vehicle is at the center of the parking place, i.e., its position is $(0.0, 0.0)$, and it has a heading of $0.0$ (this parameter ranges in the interval $[0.0, 1.0)$, representing a complete rotation). In the parking place there are 20 parking spots, 5 cars parked at lanes 3, 5, 6, 8, 13 and the target parking spot is at lane 20. The task of the DRL agent in this environment is to park the ego vehicle inside the target parking spot, with the proper heading. 

The action space of the DRL agent is composed of two actions, namely throttle and steering, both of which are continuous. The DRL agent receives a negative reward at each timestep, proportional to the Euclidean distance of the ego vehicle from the target. Moreover, it receives a constant positive reward when the target is reached and a big constant negative reward when it collides with a parked vehicle. The task is episodic with an episode finishing when either the ego vehicle is parked in the right target spot with the right heading, or it collides with a parked vehicle, or the timeout, measured in number of timesteps, expires.

We made the environment \textit{configurable}, such that the parameters of the configuration (i.e., the initial conditions at the beginning of each episode) can be changed programmatically. Correspondingly, an environment configuration needs to be \textit{valid}, i.e., it has to respect the constraints imposed by the environment. Such constraints are designed by the developers of the environment to ensure that valid configurations are \textit{solvable} by the DRL agent. In other words, any agent would be able to solve the task when starting from a valid initial configuration, since the environment does not contain any physically insurmountable obstacle or impediment.

The constraints defined for the Parking environment are the following: there cannot be a parked vehicle in the goal lane (\texttt{goal\_lane} $\notin$ \texttt{pvehicles}) since, otherwise, it would be impossible for the DRL agent to successfully park the vehicle in the target spot. Moreover, \texttt{goal\_lane} and \texttt{pvehicles} elements can vary in the interval $[1, 20]$, i.e., the target cannot be out of the parking place and there cannot be vehicles outside the parking spots. The \texttt{head\_ego} parameter can vary in the interval $[0.0, 1.0)$. The parameters \texttt{pos\_ego.x} and \texttt{pos\_ego.y} can vary in the intervals $[-10, 10]$ and $[-5, 5]$ respectively; in the former case the constraint ensures that the ego vehicle is not too far from the parking place while the latter constraint avoids the ego vehicle to be too close to parked vehicles that would make any maneuver impossible and, as a consequence, the task unsolvable.

	\section{Approach} \label{sec:approach}

The goal of our approach is to exploit the data resulting from the interaction between the DRL agent and the environment during training in order to discover the weaknesses of the DRL agent at testing time. The interaction data we consider is in the form of pairs $(\syme_i, \symc_i)$ where $\syme_i$ is the environment configuration at episode $i$ during training and $\symc_i$ is a class label, i.e., a boolean value indicating whether the DRL agent failed ($\symc_i = 1$) or not ($\symc_i = 0$) at the task in episode $i$. 

Our approach exploits the information on the failures that happened during training with the objective of generating new \textit{critical} test cases, i.e., environment configurations, in which the DRL agent under test (i.e., the DRL agent at the end of training) is likely to fail. Since the execution of the DRL agent in the environment is computationally expensive we avoid the execution of candidate new environment configurations that are not promising, i.e., that are unlikely to lead to the exposure of a failure. The training interaction data can be leveraged to predict which, among the newly generated environment configurations at testing time, are more likely to produce a failure and the DRL agent can be executed only in environments with such promising configurations.

\begin{figure*}[t]
\centering

\includegraphics[trim=0cm 14cm 2cm 0cm, clip=true, scale=0.171]
{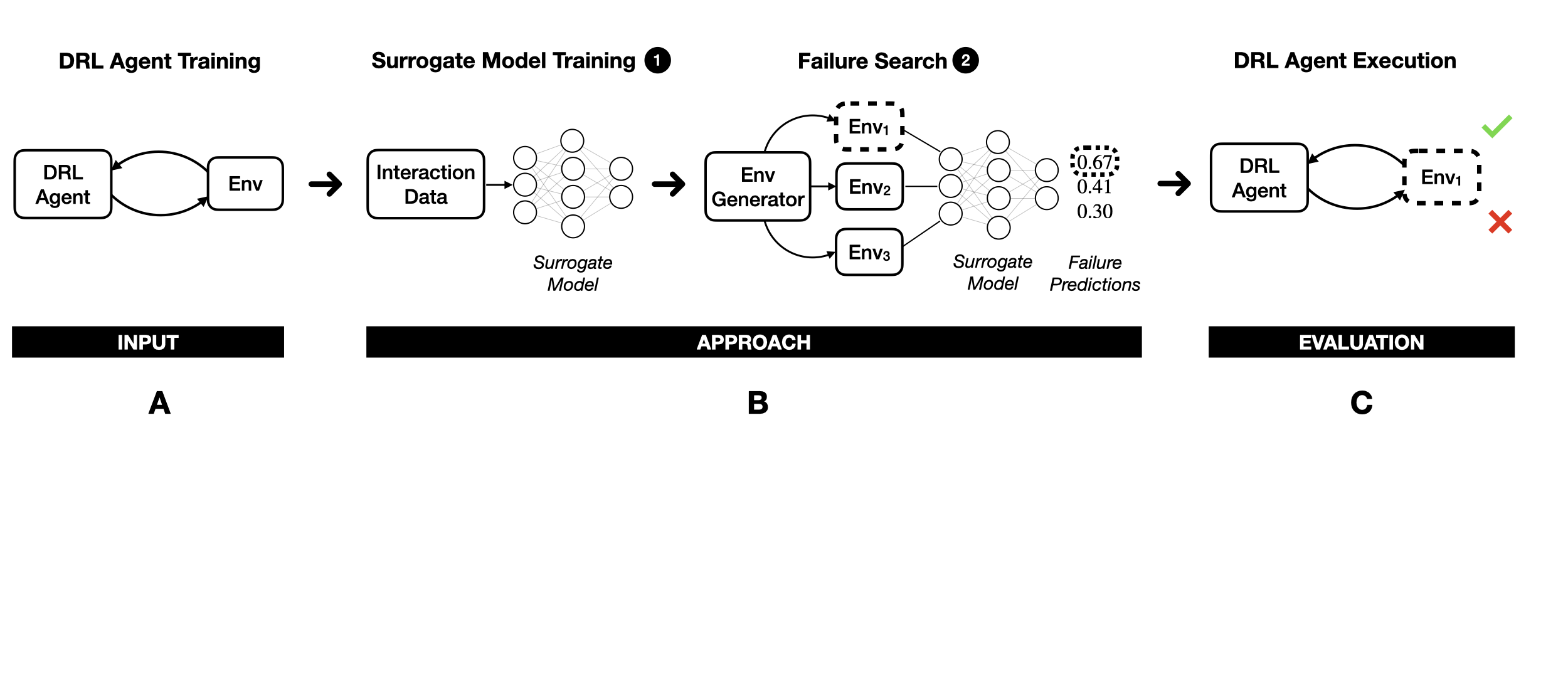}

\caption{Overall approach for testing DRL agents. Our approach takes as input a trained DRL agent that produced the interaction data (\textbf{A}) we use in the first step of our approach (\textbf{B}), i.e., \textit{Surrogate Model Training~\ding{182}}. The second step of our approach is \textit{Failure Search~\ding{183}}, that selects the environment configuration in which the DRL agent is more likely to fail. In the evaluation (\textbf{C}), we execute the selected environment configuration on the real environment to check whether the execution results in a failure or not.}
\label{fig:approach:overview} 
\end{figure*}

The current state-of-the-practice to test DRL agents is to evaluate them for a certain number of episodes, each with an environment configuration generated at random~\cite{uesato-adversarial, gym-leaderboard}. However, environment configurations generated at random are unlikely to expose failures, although specific environment configurations may exist that are challenging for the DRL agent under test.
On the other hand, the DRL training process offers a valuable source of information exploitable at testing time to efficiently expose failures of the  DRL agent under test. The intuition is that the failures experienced during training by weaker versions of the DRL agent under test are \textit{representative} of critical environment configurations of the DRL agent at the end of training, and can be used to guide the generation of new environment configurations that can challenge it.

\autoref{fig:approach:overview} summarizes the overall approach to use interaction data produced during the DRL training process to test a DRL agent. 
Our approach (i.e., \autoref{fig:approach:overview}.\textsc{B}) takes as input a trained DRL agent (i.e., \autoref{fig:approach:overview}.\textsc{A}) as well as the output of the DRL training process (i.e.,  pairs of interaction data $(\syme_i, \symc_i)$). The first step of our approach \ding{182} consists of training a surrogate model of the environment on the set of interaction data. 
The surrogate model is used in the next step (i.e., failure search) to predict whether unseen environment configurations are likely to be failures. The failure search step \ding{183} uses the surrogate model to generate environment configurations in which the DRL agent is more likely to fail, by acting as proxy for the execution of the DRL agent in the environment with such configurations. The output of step \ding{183} is the environment configuration that, considering the surrogate model a classifier, has the highest failure prediction among the candidates (in \autoref{fig:approach:overview} the selected environment configuration is \textit{Env\textsubscript{1} and it is} encircled with a dashed line). Finally, we evaluate the effectiveness of our approach by executing the DRL agent (i.e., \autoref{fig:approach:overview}.\textsc{C}) in an environment with such configuration, to check whether the execution results in an actual failure or not.

\subsection{Surrogate Model Training}

\subsubsection{Classifier}

The simplest implementation for the surrogate model is to train a classifier to predict whether, given an environment configuration, the DRL agent under test will fail in it or not. The interaction dataset $\symD = \{(\syme_1, \symc_1), \dots, (\syme_N, \symc_N)\}$ is used to train such classifier. In particular, we train a \textit{softmax} classifier (i.e., a neural network) to minimize the \textit{cross-entropy} loss. 

The classification problem to address in step \ding{182} presents some peculiar characteristics. In particular, the dataset $\symD$ might be \textit{unbalanced}, i.e., the number of environment configurations in which the agent fails ($\symc = 1$) is much lower than the number of environment configurations in which the agent does not fail ($\symc = 0$). The reason is that at the beginning of the training process the DRL agent fails in most of the environment configurations while, as training goes on, the DRL agent fails less and less until the training process converges and failures become rare. One of the strategies to train a classifier in situations of class unbalance is to introduce a weight vector $\symW$ in the loss function~\cite{torch-cross-entropy-loss-doc}. Such vector has two components, i.e., one for each class, in order to scale the loss function for each $i$-th datapoint w.r.t. the class the datapoint belongs to. The idea is to give more weight to the datapoints of the underrepresented class (i.e., the failure class) such that the classifier can learn to classify datapoints of both classes equally. In the literature~\cite{balanced-heuristic, balanced-loss-effective}, there are various proposals for the computation of $\symW$. Our implementation choice is described in \autoref{sec:evaluation:experimental-setup}.

\subsubsection{Regressor}

When the number of failures in interaction dataset $\symD$ is too small to train a reasonable classifier, another implementation choice for the surrogate model is to train a regressor. Contrary to the classifier, where it is straightforward to log when episodes are successful/unsuccessful during training, training a regressor requires defining a function that computes a continuous value for each episode. Such value should be small when the agent is close to a failure in a certain episode while it should be increasingly larger as long as the agent is far from a failure. The labels $\symc_1, \dots, \symc_N$ in the interaction dataset $\symD$ are continuous values and the training objective is to minimize the \textit{Mean Squared Error} between the prediction of the regressor and the ground-truth value in the dataset.

Similarly to the classifier, the interaction dataset $\symD$ can be unbalanced, with the majority of the data having a continuous value which represents far-from-failure situations. Also in this case, we use weighted training to cope with the unbalanced dataset (see \autoref{sec:evaluation:experimental-setup}).

In the following sections (i.e., \autoref{sec:approach:failure-prediction} and \autoref{sec:approach:failure-search}), we describe our approach considering the classifier as implementation choice for the surrogate model. Without loss of generality the same applies when the surrogate model is implemented as a regressor.

\subsection{Failure Prediction} \label{sec:approach:failure-prediction}
The classifier is used to predict the class of any environment configuration that it is not present in the dataset $\symD$ of interaction data. Therefore, it is a proxy for the actual execution of the DRL agent in the environment with a specific configuration. More formally, it is a function $\symf: \symE \rightarrow [0, 1] \in \symR$ that takes as input an environment configuration $\syme \in \symE$ and outputs a failure prediction.

Given the classifier, our objective is to generate an environment configuration that maximizes the failure prediction, i.e., to solve $\hat{\syme} = \arg \max_\syme \symf(\syme)$, and then execute the DRL agent in an environment with configuration $\hat{\syme}$. It should be noticed that the real failure prediction function $\symfs$ is only approximated by our classifier, i.e., $\symf \approx \symfs$. The reason for the approximation is two-fold: (1)~the dataset $\symD$ of interaction data might not be large and diverse enough to best represent all real-world conditions; (2)~during training the DRL agent performs non-deterministic actions to better explore the state space. As a consequence, a failure that happens during training might be due to some non-deterministic actions carried out in a specific episode. 
Despite the mismatch between $\symf$ and $\symfs$, failure search (see \autoref{fig:approach:overview}) can still effectively use the classifier that implements $\symf$, provided its feedback on the failure prediction of a new environment configuration can be reliably used to converge toward inputs that challenge the DRL agent. In fact, failure search does not require perfect guidance toward optimal inputs, it just needs a direction where the search should be directed~\cite{Jin05}.

\subsection{Search-Based Failure Search with Surrogate Models} \label{sec:approach:failure-search}

For the failure search step \ding{183} we use the output of the classifier as a \textit{fitness function} to be optimized by a search-based algorithm. In particular, we consider two search-based algorithms, i.e., \textit{Hill Climbing} and \textit{Genetic Algorithm}, to generate the environment configurations where to execute the DRL agent. 

Hill climbing is a local search algorithm that, starting from an arbitrary candidate solution to the problem (i.e., an environment configuration), incrementally changes such solution to create new ones. If a better solution is found the process is repeated until the current solution can no longer be improved or a timeout expires.
Genetic algorithm is a population-based algorithm that combines global and local search to avoid getting stuck in local optima while improving existing solutions.

\begin{algorithm}[t]
	
\DontPrintSemicolon

\SetKwInOut{Input}{Input}
\SetKwInOut{Output}{Output}
\SetKwRepeat{Repeat}{repeat}{until}

\Input{
	$f$, classifier; \\
	$NS$, neighborhood size; \\
	$e_f$, environment configuration in which the DRL agent failed during training.
}
\Output{
	$\hat{e}$, environment configuration to execute the DRL agent on.
}

\HiLi{/* Set the seed environment configuration */} \;
\eIf{$e_f$ $=$ \textit{null}}{
	$e$ $\gets$ \textsc{generateRndEnvConfig}()
}{
	$e$ $\gets$ $e_f$
}
\HiLi{/* Main loop that mutates an environment configuration $e$ guided by the classifier $f$ until the timeout expires */} \;
\Repeat{
	$\neg$ \textit{timeout()}
}{
	\HiLi{/* Assign the initial set of environment configurations and associated failure predictions */} \;
	$E$ $\gets$ $\{e\}$ \\
	$FP$ $\gets$ $\{f(e)\}$ \\
	\HiLi{/* Generate the neighborhood of the current best environment configuration */} \;
	\ForEach{$i$ $\in$ $NS$}{
		\HiLi{/* Mutate the current best environment configuration ensuring its validity */} \;
		$e_{i}$ $\gets$ \textsc{mutateEnvConfig}($e$) \\
		$fp_{i}$ $\gets$ $f(e_{i})$ \\
		$E$ $\gets$ $E$ $\cup$ $\{e_{i}\}$ \\
		$FP$ $\gets$ $FP$ $\cup$ $\{fp_{i}\}$ \\
	}
	\HiLi{/* Assign the current best environment configuration based on the results of the search */} \;
	$j$ $\gets$ $\arg \max FP$ \\
	$e$ $\gets$ $E[j]$ \\
}
$\hat{e}$ $\gets$ $e$ \\
\Return{$\hat{e}$}

\caption{Hill climbing algorithm for the generation of environment configurations}
\label{algorithm:approach:hc}
\end{algorithm}

\autoref{algorithm:approach:hc} shows the pseudocode of the hill climbing algorithm for the generation of environment configurations. It takes as input the classifier $\symf$, the size of the neighborhood $\symNS$ of the current solution $\syme$ and an optional environment configuration in which the DRL agent failed during training $\symef$. If such environment configuration is not provided, the initial solution is generated at random (see \textit{if} statement at Lines~2--6).  The \textit{for} loop at Lines~13--19 computes, at each $i$-th iteration, the \textit{neighbors} of the current solution $\syme$. Indeed, function \textsc{mutateEnvConfig} at Line~15 takes care of \textit{mutating} the current solution $\syme$ and ensuring the validity of the result, i.e., the mutation is applied to $\syme$ only if the new environment configuration $\syme_i$ is valid. The mutated solution $\syme_{i}$ is then evaluated by the classifier to compute its failure prediction at Line~16. At Lines~17--18 each mutation of the current solution, i.e., $\syme_{i}$, is stored as well as its failure prediction $\symfp_{i}$. At the end of the loop, the index $j$ of the neighboring solution with the maximum failure prediction is computed (Line~21) which is used to retrieve the corresponding neighboring solution $\syme$ (Line~22). The outermost loop at Lines~8--23 is repeated until there is search budget (i.e., the \textit{timeout}) is not expired. Finally, the best solution $\syme$ is assigned to $\hat{\syme}$ and returned.

\begin{algorithm}[t]
	
\DontPrintSemicolon

\SetKwInOut{Input}{Input}
\SetKwInOut{Output}{Output}
\SetKwRepeat{Repeat}{repeat}{until}

\Input{
	$f$, classifier; \\
	$PS$, population size; \\
	$cr$, crossover rate; \\
	$E_f$, set of environment configurations in which the DRL agent failed during training/
}
\Output{
	$\hat{e}$, environment configuration to execute the agent on.
}

\HiLi{/* Generate the initial population of environment configurations and compute the corresponding fitness */} \;
\textit{population} $\gets$ \textsc{generatePopulation}($PS$, $E_f$) \\
\textsc{computeFitness}(\textit{population}, $f$) \\
\textit{currentIteration} $\gets$ $0$ \\
\HiLi{/* Main loop that changes the population guided by the classifier $f$ until the timeout expires */} \;
\Repeat{
	$\neg$ \textit{timeout}()
}{
	\HiLi{/* Build the new population by extracting a certain percentage of the best environment configurations */} \;
	\textit{newPop} $\gets$ \textsc{elitism}(\textit{population}) \\
	\HiLi{/* Fill the rest of the population by evolving the environment configurations */} \;
	\While{$|$ \textit{newPop} $|$ $<$ $PS$}{
		\HiLi{/* Select the best environment configurations according to their fitness */} \;
		$pe_1$ $\gets$ \textsc{selection}(\textit{population}) \\
		$pe_2$ $\gets$ \textsc{selection}(\textit{population}) \\
		
		\HiLi{/* Copy the environment configurations (offspring) to avoid changing the original ones (parents) */} \;
		$oe_1$ $\gets$ \textsc{copy}($pe_1$) \\
		$oe_2$ $\gets$ \textsc{copy}($pe_1$) \\
		
		\HiLi{/* Crossover two environment configurations with a certain probability $cr$, ensuring their validity */} \;
		\If{\textsc{getRandomFloat}() $<$ $cr$}{
			$oe_1$, $oe_2$ $\gets$ \textsc{crossover}($oe_1$, $oe_2$)
		}
		
		\HiLi{/* Mutate the offsprings ensuring their validity */} \;
		$oe_1$ $\gets$ \textsc{mutateEnvConfig}($oe_1$) \\
		$oe_2$ $\gets$ \textsc{mutateEnvConfig}($oe_2$) \\
		
		\HiLi{/* Add the best environment configurations to the population according to their fitness */} \;
		\textsc{addBestIndividuals}(\textit{newPop}, $pe_1$, $pe_2$, $oe_1$, $oe_2$) \\

	}
	\HiLi{/* Compute the fitness of environment configurations in the new population */} \;
	\textit{population} $\gets$ \textit{newPop} \\
	\textsc{computeFitness}(\textit{population}, $f$) \\
	\HiLi{/* Replace the worst environment configurations in the population to avoid stagnation */} \;
	\textsc{reseedPopulation}(\textit{population}, \textit{currentIteration}, $E_f$) \\
	\textit{currentIteration} $\gets$ \textit{currentIteration} $+ 1$ \\
}
\HiLi{/* Extract the environment configuration with the best fitness */} \;
$\hat{e}$ $\gets$ \textsc{getIndividualWithBestFitness}(\textit{population}, $f$) \\
\Return{$\hat{e}$}

\caption{Genetic algorithm for the generation of environment configurations}
\label{algorithm:approach:ga}
\end{algorithm}

The \textit{Genetic Algorithm} shown in \autoref{algorithm:approach:ga} takes as input the classifier $\symf$, the population size $\symPS$, the crossover rate $\symcr$ and an optional set of environment configurations in which the DRL agent failed during training $\symEf$. Such set can be used to fill the initial population when it is available; otherwise the initial population is generated randomly (Line~2). The \textsc{computeFitness} procedure at Line~3 computes the fitness value for each solution in the population, which is its failure prediction as computed by the classifier $\symf$. At Line~8 a new population with the best solutions from the current population is instantiated (\textit{elitism}). The while loop at Lines~10--26 is the evolution part of the algorithm which terminates when the size of the new population reaches the target population size $\symPS$. At the beginning of the loop two solutions are selected based on their fitness (Lines~12--13) and are crossed over according to the crossover probability $\symcr$ (Lines~18--20); the solutions $\symoe_1$ and $\symoe_2$ are modified only if the crossover results are two valid configurations. Afterward, the solutions are mutated (Lines~22--23, also in this case $\symoe_1$ and $\symoe_2$ are modified only if the changes result in valid configurations) and finally the best solutions (either the parents $\sympe_1$, $\sympe_2$ or the offsprings $\symoe_1$, $\symoe_2$) are stored in the new population (Line~25). At the end of the while loop the current population is assigned the new population and the fitness value of each solution is computed (Lines~28--29). The \textsc{reseedPopulation} procedure at Line~31 takes care of avoiding stagnation by reseeding the population according to the \textit{currentIteration} variable. If the set of failure environment configurations $\symEf$ is available,  the solutions with the worst fitness in the current population are replaced by randomly sampled solutions from the set $\symEf$; otherwise the worst solutions are replaced by randomly generated individuals. The main loop (Lines~6--33) terminates when the search budget expires. Finally, the solution with the best fitness $\hat{\syme}$ is taken at Line~35 and returned.

Both search algorithms support seeding from existing environment configurations that caused a failure of the agent during training. In particular, let us suppose that the parameter $\symef$ is not \textit{null} in \autoref{algorithm:approach:hc} (similarly, $\symEf$ is not empty in \autoref{algorithm:approach:ga}). The DRL agent may not fail in such environment configurations because it may have encountered similar environment configurations later during training, and it might have adapted to them. However, despite adaptation, the DRL agent might not perform well and a \textit{proper change} in such environment configuration has the potential to trigger a failure. For example, let us suppose that the environment configuration $\symef = [20, 0.0, \{3, 5, 6, 8, 13, 19\}, (0.0, 0.0)]$ causes a failure of the DRL agent during training in the Parking environment (it is the same environment configuration shown in \autoref{fig:background:parking-environment}, except that there is a parked vehicle beside the target spot), but the DRL agent does not fail in $\symef$ at testing time. However, changing the heading of the ego vehicle from 0.0 to 0.5, i.e., the opposite direction w.r.t. the target spot, might result in the DRL agent not being able to turn the vehicle and park it with the right heading, since the mutation has succeeded in making the environment more challenging, eventually exposing a failure.

\subsubsection{Mutation Function}

Knowing what to change in the environment configuration is important for finding failures. The function that makes this decision is the \textsc{mutateEnvConfig} function (Line~13 in \autoref{algorithm:approach:hc} and Lines~14--15 in \autoref{algorithm:approach:ga}). %
We propose two implementations of the \textsc{mutateEnvConfig} function: the first one randomly changes a parameter of the given environment configuration, choosing among all parameters with equal probability. The second one only changes the parameter of an environment configuration that contributes the most to the failure prediction. The idea is that some parameters are more \textit{critical} than others when considering how they affect failure prediction. Hence, changing them is more beneficial for generating failure environment configurations than changing the other parameters. In particular, we propose to use the \textit{saliency} method to compute  \textit{input attribution}~\cite{saliency-paper}, i.e., to determine how much an input influences a prediction made by a neural network (our classifier). Given an environment configuration and the classifier, the saliency method computes the input gradient, i.e., the partial derivatives over all the individual parameters of the environment configuration. The absolute value of each gradient indicates which input parameter is more critical to the failure prediction and its sign indicates the direction (i.e., positive or negative) of change. In our saliency-guided implementation of the \textsc{mutateEnvConfig} function we take the output computed by saliency and change the parameter in the environment configuration whose corresponding gradient is the highest one; the direction of change, either positive or negative, depends on the sign of the gradient. 

For instance, let us consider the environment configuration: $\syme = [20, 0.0, \{3, 5, 6, 8, 13, 19\}, (0.0, 0.0)]$, i.e., an environment configuration of the Parking environment. The input attribution for this environment configuration is an array of the same size as the input, as it contains the partial derivatives over each input. Let us suppose that the highest value in the array is in second position, i.e., the position corresponding to the parameter \texttt{head\_ego}, and that such value is positive. This means that the parameter \texttt{head\_ego} is the most critical one affecting the failure prediction of the classifier and that changing it in the positive direction, i.e., increasing it, will also increase the failure prediction for the resulting environment configuration. 

\subsubsection{Crossover Function} \label{sec:approach:crossover}

The crossover function is specific to the genetic algorithm (Line~12 in \autoref{algorithm:approach:ga}). We propose a single-point crossover implementation where, given two environment configurations, the cut point is determined randomly; then the elements in the two configurations before and after the cut point are swapped. We chose this simple implementation for  crossover because it can be applied to configurations of any case study with little to no modifications. Custom implementations that take into account the specific features of each case study remain possible in our implementation.

For instance, let us suppose that $\syme_1 = [20, 0.0, \{3, 5, 6, 8, 13, 19\}, (0.0, 0.0)]$ and $\syme_2 = [15, 0.5, \{1, 3, 9\},(-1.0, 7.5)]$ are two Parking environment configurations. The cut point is computed based on the number of parameters in the environment configuration, which in the case of Parking is equal to four (i.e., \texttt{goal\_lane}, \texttt{head\_ego}, \texttt{pvehicles} and \texttt{pos\_ego}). Let us suppose that the cut point is at first position. The two environment configurations after crossover will be as follows: $\symce_1 = [20, 0.5, \{1, 3, 9\},(-1.0, 7.5)]$ and $\symce_2 = [15, 0.0, \{3, 5, 6, 8, 13, 19\}, (0.0, 0.0)]$.

\subsection{Implementation}

We implemented our approach in a Python tool called \tool (Latin for ``search'') which is publicly available~\cite{replication-package}. The DRL agents are implemented by the \textit{stable-baselines}~\cite{sb3-paper} library and we use \textit{Pytorch}~\cite{pytorch} and \textit{scikit-learn}~\cite{scikit-learn} to implement the training and inference of the classifier. The \textit{saliency} method is implemented by the \textit{captum} library~\cite{captum}.

	\section{Empirical Evaluation} \label{sec:evaluation}

We consider the following research questions: 

\noindent
\textbf{RQ\textsubscript{1} (Effectiveness):}
\textit{What is the effectiveness of the proposed approach? Can it generate failure environment configurations for the DRL agent under test?}

\noindent
\textbf{RQ\textsubscript{2} (Comparison):}
\textit{How does \tool compare against the random baseline? How does it compare against the state-of-the-art sampling approach?}

\noindent
\textbf{RQ\textsubscript{3} (Hyperparameters):}
\textit{What is the impact of the key hyperparameters of \tool?}

\noindent
\textbf{RQ\textsubscript{4} (Ineffective Failure Seeds):}
\textit{How does the effectiveness of \tool change when failure seeds are ineffective at testing time?}

RQ\textsubscript{1} evaluates the effectiveness of \tool, i.e., its capability to generate failure environment configurations for the DRL agent under test. This research question acts as a sanity check to make sure that our approach is able to generate failures that we can analyze and study in the next research question.

RQ\textsubscript{2} compares \tool with the random baseline and the state-of-the-art sampling approach~\cite{uesato-adversarial}, both w.r.t. the number of generated failure environment configurations and their diversity. %

RQ\textsubscript{3} investigates the key hyperparameters of \tool both w.r.t. the number of failure environment configurations that are generated by each hyperparameter setting and their diversity. We first analyze the choice of the search algorithm, i.e., hill climbing vs genetic algorithm. Second, since our approach can work both with and without the provisioning of failure seeds ($\symef$ and $\symEf$ are optional parameters in \autoref{algorithm:approach:hc} and \autoref{algorithm:approach:ga}, respectively), we want to investigate which of the two seed strategies is more convenient to use. Third, when mutating the environment configurations \tool can either choose them randomly or it can focus on those that have the highest influence on the failure prediction, as determined by the saliency method. Hence, we want to evaluate the impact of the mutation strategy when using \tool.

In RQ\textsubscript{4} we study how \tool performs when failure seeds, i.e., failure configurations causing a failure of the agent during training, are not effective for the DRL agent at testing time. In this research question we want to study whether \tool is still effective at finding failures when there is no guidance from the failure seeds.

\subsection{Case Studies} \label{sec:evaluation:case-studies}

\head{Parking} We already introduced the first environment, i.e., Parking~\cite{highwayenv}, in \autoref{sec:background}, where we also describe the representation of the environment configuration (see \autoref{fig:background:parking-environment}). Such environment has been used in several studies, especially to evaluate the capabilities of DRL agents~\cite{highway-env-robust-predictable-control,highway-env-accelerated,highway-env-corner,highway-env-deep,highway-env-intelligent,highway-env-learning-interaction,highway-env-quick}.

The encoding adopted to train a classifier for failure prediction consists of an array of 24 elements where the first two values are for the two single-value parameters of the environment configuration (i.e., \texttt{goal\_lane} and \texttt{head\_ego}), followed by 20 values corresponding to the one-hot encoding of the \texttt{pvehicles} parameter (i.e., each position has a value of 1 only when there is a vehicle in the respective spot; a value of 0 otherwise) and finally the two values of the \texttt{pos\_ego} parameter.

Regarding the mutation operators, the \texttt{goal\_lane} parameter value is increased or decreased, with equal probability, and the change amount is an integer in the interval $[1, 20]$. The \texttt{heading\_ego} parameter value is increased or decreased, with equal probability, and the change amount is a random floating point number in the interval $[0, 1)$. The coordinates of the \texttt{pos\_ego} parameter tuple are increased or decreased, with equal probability, by a random float in the interval $[0, 1]$. For what concerns the \texttt{pvehicles} parameter, with equal probability parked cars in the \texttt{pvehicles} set are added/removed or the parked car occupancy indexes are mutated. In the former case a certain number of indices is selected at random for adding or removing parked cars with equal probability. 
In the latter case a certain number of parked car indices is selected at random  to be mutated, i.e., increased or decreased by a random integer  in the interval $[1, 20]$. Moreover, the crossover operator considers the parameters of the Parking environment configurations as unique features. An example of crossover for the Parking environment is in \autoref{sec:approach:crossover}. 

\head{Humanoid} The second environment we consider is \textit{Humanoid}, built using the \textit{Mujoco} simulator~\cite{mujoco}. Mujoco is a very popular simulator in the DRL community, especially to benchmark DRL algorithms in continuous control tasks. The 3D physics simulator has different pre-built environments, with Humanoid being one of the most challenging environments for DRL algorithms~\cite{humanoid-challenging}. %

\begin{figure}[t]
\centering

\includegraphics[trim=0cm 22cm 21cm 0cm, clip=true, scale=0.235]
{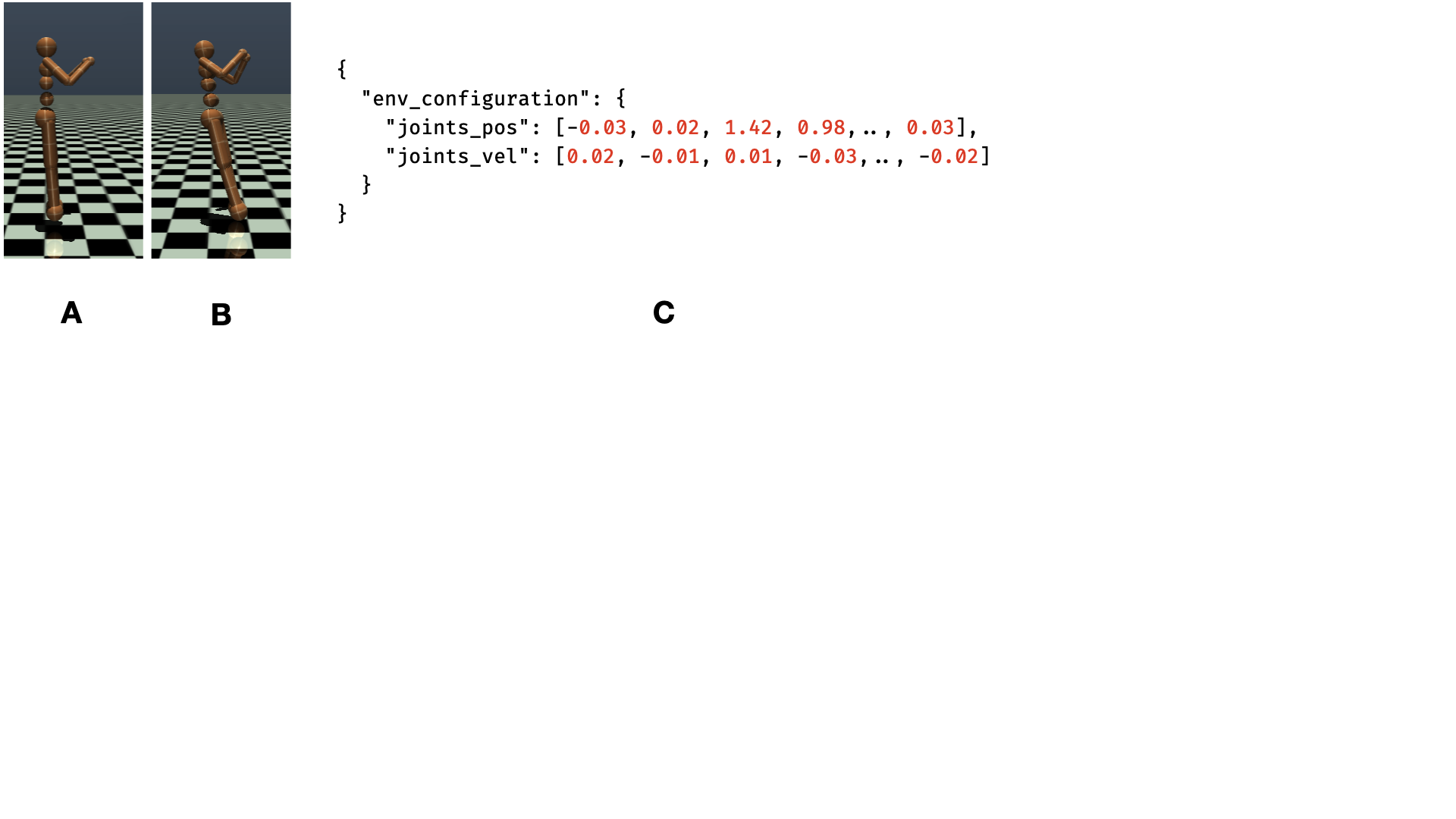}

\caption{Two initial configurations (\textbf{A} and \textbf{B}) of the \textit{Humanoid} environment in the \textit{Mujoco} simulator~\cite{mujoco}. The right-hand side (\textbf{C}), shows an example of environment configuration.} 
\label{fig:empirical-evaluation:humanoid-environment} 
\end{figure}

In Humanoid the DRL agent needs to control a bipedal robot in a 3D space and make it walk on a smooth surface. Each environment configuration is composed of two arrays, i.e., \texttt{joints\_pos} and \texttt{joints\_vel}. The former has size 24 and consists of the positions and rotations of the joints of the robot while the latter has size 23 and consists of the linear and angular velocities of those joints (more information are available in the Gym Library wiki~\cite{humanoid-wiki}). All the angles are in radians, represented as floating point numbers. \autoref{fig:empirical-evaluation:humanoid-environment} shows two configurations of the Humanoid environment; we can see that the initial joints positions can be altered, forming different initial configurations for the Humanoid robot (\autoref{fig:empirical-evaluation:humanoid-environment}.\textbf{A} and \autoref{fig:empirical-evaluation:humanoid-environment}.\textbf{B} show the rendering of two different environment configurations, while \autoref{fig:empirical-evaluation:humanoid-environment}.\textbf{C} shows how an environment configuration is encoded.). %

The state space of the DRL agent is an array of size 376, composed of joint positions, angles and  relative velocities, plus other components as center of mass inertia and velocity. The action space is composed of 17 elements that are the degrees of freedom of the robot, i.e., the actuated joints of the robot. The reward function encourages the DRL agent to walk as fast as possible plus a bonus for each timestep. An episode terminates when the abdomen $y$ coordinate of the robot goes out of the range $(1, 2)$, which indicates that the robot has fallen down, or when a timeout expires (we set such timeout to  300 timesteps). When the robot \textit{falls} we deem the episode unsuccessful; on the other hand, when the timeout expires we consider the episode successful. We changed the environment interface to be \textit{configurable} such that the parameters \texttt{joints\_pos} and \texttt{joints\_vel} can be set programmatically at the beginning of each episode. The original implementation initializes \texttt{joints\_pos} to $[0, 0, 1.4, 1, 0, \dots, 0]$ and \texttt{joints\_vel} to all zeros; then, to generate a new environment configuration, each value in both arrays can be changed by adding or subtracting a small quantity $\symm$ (which is set to 0.03). We followed the original implementation to define \textit{validity}, i.e., we consider an environment configuration to be valid only if the values of its parameters, i.e., \texttt{joints\_pos} and \texttt{joints\_vel}, are within the interval $[-\symm, \symm]$ w.r.t. the initial values of such parameters. For instance, a valid environment configuration can have the third value of its \texttt{joints\_pos} array within the interval $[1.4 - \symm, 1.4 + \symm]$, since that value is initialized to 1.4.

The encoding we adopt as input to the classifier is the concatenation of the two arrays that define the environment configuration, i.e., \texttt{joints\_pos} and \texttt{joints\_vel}. Regarding the mutation operators, they are defined in the same way for both parameters, i.e., once an index of either \texttt{joints\_pos} or \texttt{joints\_vel} is chosen at random, the value at that index is either decreased or increased, with equal probability, by a random floating point quantity in the interval $[-\symm, \symm]$. Moreover, the crossover operator is based on the parameters of the Humanoid environment, as in Parking. In Humanoid there are only two parameters, i.e., \texttt{joints\_pos} and \texttt{joints\_vel}; therefore, during crossover, the entire \texttt{joints\_pos} and \texttt{joints\_vel} vectors are swapped between two different environment configurations.

\head{Driving} The third environment we consider is \textit{Driving}, built using the \textit{DonkeyCar} simulator~\cite{sdsandbox}. This platform has been used in previous works to train and test self-driving car software both based on supervised learning and reinforcement learning~\cite{donkey-drl-autonomous-driving,viitala,donkey-scaled-vehicles,donkey-drl-self-driving-scale}.

\begin{figure}[t]
\centering

\includegraphics[trim=0cm 20cm 4cm 0cm, clip=true, scale=0.23]
{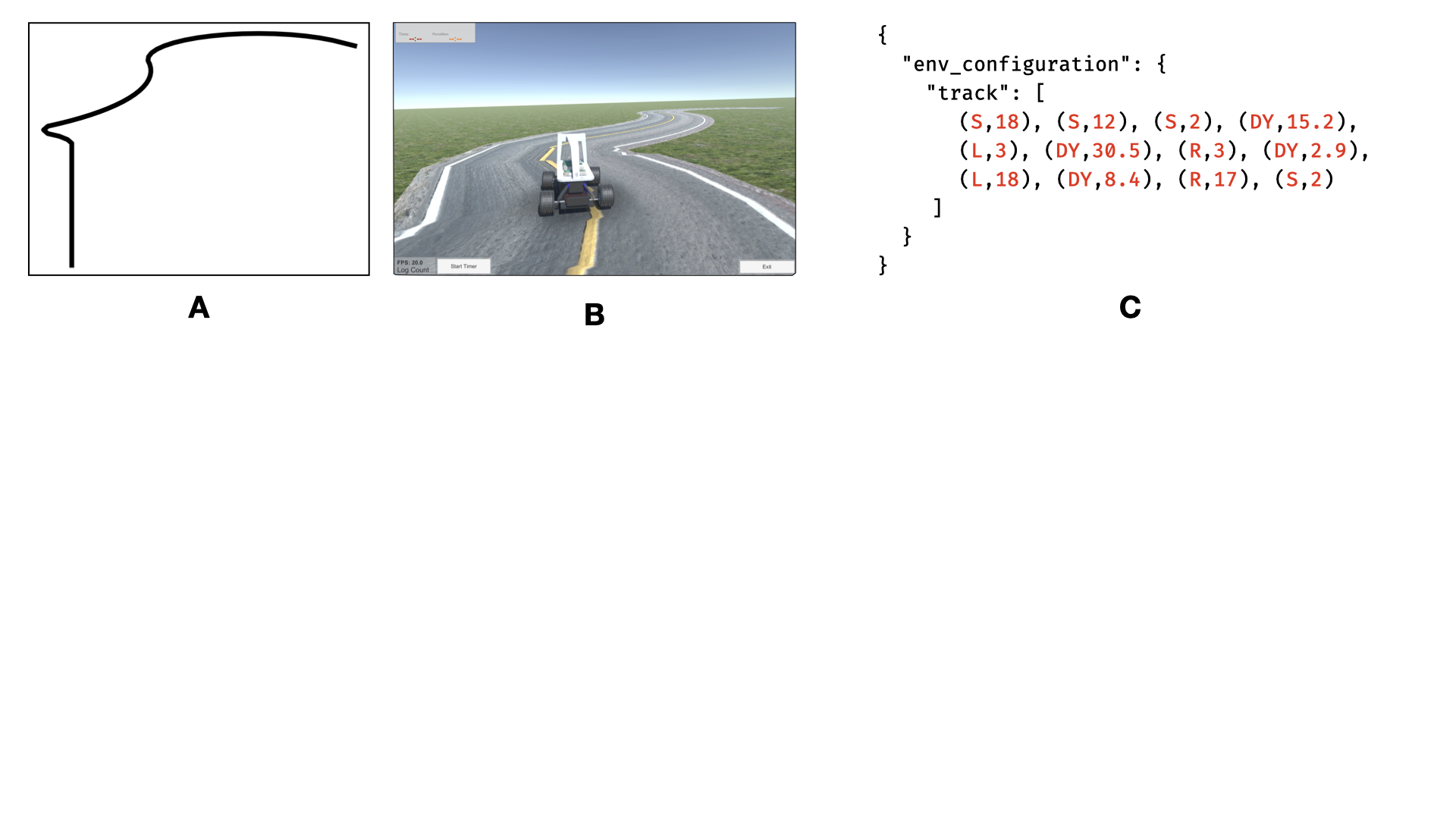}

\caption{An initial configuration of the \textit{Driving} environment in the \textit{DonkeyCar} simulator~\cite{sdsandbox}. The left-hand side shows both the plot (\textbf{A}) of the environment configuration on the right-hand side (\textbf{C}), and how the configuration is rendered in the \textit{DonkeyCar} simulator (\textbf{B}).} 
\label{fig:empirical-evaluation:driving-environment} 
\end{figure}

In this environment the DRL agent controls a car which drives along a track. The configuration determines the shape of the track in which the car drives (see \autoref{fig:empirical-evaluation:driving-environment}.\textbf{A}). The track is represented as a list of 12 pairs, where each pair consists of two elements, i.e., a command and a value (see \autoref{fig:empirical-evaluation:driving-environment}.\textbf{C}). The possible commands are \texttt{S} which indicates a straight line, \texttt{R} which indicates a right curve, \texttt{L} which indicates a left curve and \texttt{DY} which signals the beginning of a curve. The value associated to each command represents the number of road units, except for the \texttt{DY} command, where the associated value represents the individual angle of rotation for each road unit. For example, the sequence of commands \texttt{[(S,2), (DY,15.2), (L,3)]} (i.e., the first curve of the track represented by the environment configuration in \autoref{fig:empirical-evaluation:driving-environment}.\textbf{B}) instructs the road engine to build a straight road for two road units followed by a left curve which is three road units long, for a total of 45.6 degrees (i.e., $3 \times 15.2$). 

The state space of the DRL agent is an RGB image of size (160, 120) taken by the front camera of the car, while the action space is two-dimensional, i.e., steering angle and throttle. The DRL  agent receives a small negative reward per timestep plus a positive reward every time it crosses a waypoint in the track, minus a penalty related to the \textit{cross track error}, i.e., the distance between the center of mass of the car and the center of the track. The first component of the reward encourages the DRL agent to go as fast as possible, the second privileges the progress on the track and the third forces the DRL agent to drive as close to the center of the track as possible. Moreover, the DRL agent receives a large negative reward when it goes off-road. In such cases the episode terminates unsuccessfully, whereas when the DRL agent crosses the end line of the track, the episode is deemed successful. 

We modified the \textit{DonkeyCar} simulator to make it \textit{configurable}, such that at the beginning of each episode the track passed as input can be instantiated in the simulator. Regarding \textit{validity}, there are a number of constraints that need to be respected in order for the track to be valid. First, each track should start and end with an \texttt{S} command. Moreover, after a \texttt{DY} command there must be either an \texttt{L} or \texttt{R} command. 
Afterward, the track must not contain loops (i.e., self-intersections), it must not have very sharp turns (i.e., with a rotation angle $> 170\degree$) and it must have at least 3 curves, one of which must be with a rotation angle of at least $120\degree$. The last two constraints ensures that the generated tracks are non-trivial.

The encoding we adopt as input to the classifier is the concatenation of two arrays, i.e., the array of all commands, where each command is given a unique integer identifier, and the array of all values. Regarding the mutation operators, we define one for commands and one for values and, when analyzing a command-value pair, we either change the command or the value with equal probability. The change command operator can only change an \texttt{L} command to an \texttt{R} command or vice-versa. The change value operator first analyzes the associated command and, if it is a \texttt{DY} command, it either increases or decreases, with equal probability, the current value by a random floating point number in the interval $[0, 50]$. Otherwise, it increases or decreases, with equal probability, the current value by a random integer number in the interval $[1, 20]$. 
Moreover, the crossover operator considers the Driving configuration as a list of command-value pairs. The cut point is a value in the interval $[1, 12]$ such that after crossover the resulting Driving environment configurations have command-value pairs coming from both parent environment configurations. We customized the crossover operator for this environment by retrying crossover a certain number of times, until either both resulting configurations are valid or the maximum counter is reached. 

\begin{table}[ht]
\setlength{\tabcolsep}{2.5pt}
\renewcommand{\arraystretch}{1.2}
\centering
\caption{Case studies training metrics}
\begin{tabular}{rccccccc}
\toprule

\multicolumn{1}{l}{} & 
\multicolumn{1}{l}{} & 
\multicolumn{3}{c}{RQ\textsubscript{1}---RQ\textsubscript{3}} & 
\multicolumn{3}{c}{RQ\textsubscript{4}} \\

\cmidrule(r){3-5} 
\cmidrule(r){6-8}

\multicolumn{1}{l}{} & 
DRL algo & 
\rot{\# tr. timesteps} & 
\rot{\# tr. episodes} & 
\rot{\# tr. failures} & 
\rot{\# tr. timesteps} & 
\rot{\# tr. episodes} & 
\rot{\# tr. failures} \\

\midrule

\textsc{Parking} & HER~\cite{her} + TQC~\cite{tqc} & 200\textit{k} & 8789 & 206 & 80\textit{k} & 3159 & 17 \\
\textsc{Humanoid} & TQC~\cite{tqc} & 1500\textit{k} & 6965 & 270 & 300\textit{k} & 1141 & 318 \\
\textsc{Driving} & SAC~\cite{sac} & 1000\textit{k} & 11274 & 194 & 500\textit{k} & 1445 & 13 \\

\bottomrule
\end{tabular}
\label{table:empirical-evaluation:case-studies}
\end{table}

\subsection{Experimental Setup} \label{sec:evaluation:experimental-setup}

\subsubsection{Procedure} 

We trained the DRL agents in each environment using the hyperparameters recommended by Raffin et al.~\cite{sb3-paper, rl-zoo3}. For RQ\textsubscript{4} we simplified the training environments in order to make the resulting agents more robust against training failure configurations. In particular, we removed the parked vehicles in Parking, we decreased the variation of the two parameter vectors determining the initial configurations in Humanoid (i.e., we set $\symm$ to 0.01 instead of 0.03, see the respective paragraph in \autoref{sec:evaluation:case-studies}), and we constrained the number of curves in Driving to be $\ge 2$ instead of $\ge 3$, of which the hardest one must be $\le 130\degree$ instead of $\le 170\degree$.
\autoref{table:empirical-evaluation:case-studies} shows in Column~1 the algorithms we used for each environment. Column~2 (resp. Column~5 for RQ\textsubscript{4}) shows the number of training timesteps which corresponds to the number of episodes shown in Column~3 (resp. Column~6 for RQ\textsubscript{4}). In each setting we trained each DRL agent until convergence, i.e., until the success rate in the last 100 episodes was 100\%.  Column~4 and Column~7 show the number of training failures in the two settings after filtering the initial 30\% of the data. We can observe that, in the first setting (i.e., Column~4), the number of failures remaining after filtering is comparable and close to 200 on average. In the second setting (i.e., Column~7), the number of failures is much lower in Parking and Driving (i.e., $\approx$20 \textit{vs} $\approx$200), while in Humanoid the DRL agent experiences a drop in performance after the initial training phase in both settings, making the number of training failures after filtering comparable.

Given the low number of failures in the second setting, we resort to a regressor-based surrogate model. Indeed, a regressor, as opposed to a classifier, can learn from near-failure interactions during training, i.e., where the agent got close to failing but ultimately such interactions were successful. For the Parking environment we considered the length of the episode as continuous value. In Parking, an episode finishes unsuccessfully when the agent is not able to park the vehicle in a certain amount of time. A long episode indicates a challenging environment configuration where the agent struggled to complete the parking task.

In Humanoid, we consider the abdomen latitudinal coordinate of the robot, which determines when the robot has fallen down; for a given episode we consider the minimum distance between the current coordinate and the respective lower and upper bounds. In Driving, we use the maximum cross track error in an episode as a continuous value for the regressor.

In all case studies, we normalize the continuous values between 0 and 1, where 0 indicates a failure. When using the regressor-based surrogate model for failure search, our objective is to generate environment configurations that minimize the prediction of the regressor.

\head{Classifier Training} To compensate for the unbalanced dataset when training the failure classifier, we compute a weight vector $\symW$ as follows: given $\symN$ datapoints, the array of class targets $\symY$, with $\symCbar$ being the number of classes: 

\begin{equation}
	\symW = \frac{\symN}{\symCbar \cdot \texttt{hist}(\symY)}
\end{equation} 
\noindent
where \texttt{hist($Y$)} is a function that outputs an array of size $\symCbar$ indicating the number of datapoints for each class. Such formula, gives a higher weight to the underrepresented classes~\cite{balanced-heuristic}.

We chose a multi-layer perceptron as the classifier architecture. The reason is twofold: first, the size of the available training data is small (at most 10\textit{k} examples for both failure and non-failure  classes, see \autoref{table:empirical-evaluation:case-studies}). Secondly,  our environment configurations are small size one-dimensional feature vectors (24 for Parking, 47 for Humanoid and 24 for Driving). Furthermore, since the output of the classifier is used as a fitness function during testing, adopting complex models to process simple inputs would be inefficient and prone to overfitting. 

When training the failure predictor, there are two important hyperparameters to consider. The first one is the amount of initial interaction data to filter out. Indeed, at the beginning of training the agent carries out random actions and, as a consequence, it often fails regardless of the environment configuration. This means that the \textit{earliest} failures are not useful to predict the failures of the DRL agent under test. The other hyperparameter that depends on the case study is the number of hidden layers of the multi-layer perceptron. 

In our experiments we considered nine levels of filtering (i.e., 5, 10--80), where, for instance, 5\% filtering means that the first 5\% of the environment configurations are not considered for training the classifier. Moreover, we chose four different number of hidden layers, i.e., from 1 to 4, where each hidden layer is composed of 32 units followed by a batch normalization layer~\cite{batchnorm} and a dropout layer~\cite{dropout} with probability of dropout of 0.5 (except when the network has only one hidden layer). For each pair filter-layers we trained the classifier ten times (with all the other training hyperparameters, e.g., learning rate, fixed), every time with a different random seed. We used a validation set formed using 20\% of the data to save the best model during training which we evaluated on a held-out test set. We built such test set by choosing a filtering level of 5\% for each case study and by selecting 10\% of the data at random. 
During hyperparameter tuning we measured precision and recall of the classifier considering the failure class as the positive class. 
We deem precision more important than recall since in order to guide failure search it is more important for the classifier to be as precise as possible (i.e., few false positives) even at the cost of missing some failures (i.e., false negatives).
We chose the best classifier model by looking at the results of the ten training runs for each filter-layers pair, and we selected the model with the highest precision and with a recall of at least 10\%, in order not to miss too many failures.

The best classifier for \textit{Parking} has four hidden layers and a filtering level of 50\%, reaching a precision of 24\% and a recall of 17\% on the held-out test set. The best classifier for Humanoid has one hidden layer and a filtering level of 10\%, reaching a precision of 62\% and a recall of 48\% on the held-out test set. Finally, the best classifier for Driving has four hidden layers and a filtering level of 30\%, reaching a precision of 25\% and a recall of 12\% on the held-out test set. Such classifiers are used in the failure search step for each case study.

\head{Regressor Training} In order to deal with the unbalanced regression problem we use the Label Distribution Smoothing (LDS) approach proposed by Yuzhe et al.~\cite{delving-regression}. The approach starts by convolving a symmetric kernel with the empirical distribution of continuous values (e.g., obtained by binning the continuous values). The output is a smoothed version of the distribution that we use in the loss function to proportionally weight each continuous value based on their frequency.

The best regressor for \textit{Parking} has two hidden layers and a filtering level of 10\%; the best regressor for Humanoid has two hidden layers and a filtering level of 30\%; the best regressor for Driving has two hidden layers and a filtering level of 5\%. Such regressors are used in the failure search step for each case study when training failures are ineffective at testing time and correspondingly a regressor is potentially a better surrogate model than a classifier.

\head{Baselines} We compared \tool with two approaches. The first approach is the \textit{random} baseline, where environment configurations are generated at random. Such baseline is useful to understand whether the proposed approach is able to outperform the state-of-the-practice in testing DRL agents~\cite{uesato-adversarial,gym-leaderboard}.
The second approach is the state-of-the-art \textit{sampling} approach by Uesato et al.~\cite{uesato-adversarial}. This simple approach consists of generating a large initial set of $\symT$ environment configurations at random and choosing the one that, according to the classifier $\symf$, has the highest failure prediction. %

\head{\tool} We considered hill climbing and genetic algorithm each with four different settings. In the first one, the seed environment configurations to evolve are generated at random (hc\textsubscript{rnd} and ga\textsubscript{rnd}). In the second, we used hill climbing and genetic algorithm to evolve failure environment configurations (hc\textsubscript{fail} and ga\textsubscript{fail}). Regarding the settings with the failure seeds, we always filtered out the initial 30\% of the environment configurations, in order not to include failures that are likely not representative of the failures of the DRL agent under test. Furthermore, we considered hill climbing and genetic algorithm guided by saliency, evolving both random  (hc\textsubscript{sal+rnd} and ga\textsubscript{sal+rnd}) and failure (hc\textsubscript{sal+fail} and ga\textsubscript{sal+fail}) environment configurations. For \tool and the sampling approach we considered a fixed search budget $\symB$ per environment configuration. In particular, we chose $B = 5$ seconds for Parking and Humanoid, while $B = 30$ seconds for Driving, as from preliminary experiments we observed that such budget was enough to reach fitness convergence for all approaches.

During failure search, we generated $\symT = 100$ environment configurations for each case study and each approach, and we evaluated the respective DRL agent in each environment configuration. 
Since Humanoid and Driving simulators are non-deterministic we evaluated the respective DRL agents in each environment configuration ten times. Moreover, we repeated the experiments ten times for each failure search approach in order to cope with the intrinsic randomness of the approaches. Overall, we have 3 case studies, 10 techniques (8 settings of \tool, including hill climbing vs genetic algorithm, random vs failure seed, two mutation strategies (i.e., random and saliency), plus the random and the sampling approaches), each generating 100 environment configurations (each repeated 10 times in Humanoid and Driving). Finally, each technique is executed 10 times for a total of 210\textit{k} simulations.

\head{Hardware Resources} Due to the high number of simulations, we resorted to the university cluster with 20 CPU nodes, and parallelized the execution of the experiments to obtain the results in a reasonable amount of time. We did not make use of GPU nodes in the experiments, as DRL models are typically small and simulators could be executed headless on CPUs.

\subsubsection{Metrics.} \label{sec:evaluation:metrics}

In order to assess the \textit{effectiveness} of our failure search approach (RQ\textsubscript{1}) and to compare it with the baselines (RQ\textsubscript{2}), we measured the number of failures each approach triggers given the number of environment configurations to generate (i.e., $\symT$) and the fixed search budget $\symB$ per environment configuration. In non-deterministic environments, as Humanoid and Driving, we measured the failure probability of each environment configuration out of ten runs and we considered an environment configuration to cause a failure if its failure probability is $> 0.5$. As each failure search approach is executed ten times, we determined whether there is a statistical difference between the failures triggered by each pair of failure search approaches (including the 8 settings of \tool) by computing the Mann-Whitney U Test~\cite{mann-whitney-u-test, arcuri-hitchhiker}. To measure the effect size, we computed the Vargha Delaney metric $\symVD$~\cite{vargha-delaney}.

To compare the competing approaches (RQ\textsubscript{2}) we also considered two types of diversity regarding the failures generated by each approach, namely input and output diversity. For \textit{input diversity}, we first clustered all the environment configurations that caused a failure, across all considered failure search techniques, obtaining a single partition of all such failing environment configurations. Similarly, for \textit{output diversity}, we clustered the \textit{trajectories} (i.e., positions over time) of the DRL agents on the failure inducing environment configurations. In Parking and Driving we considered the trajectories of the vehicle while performing the task, i.e., parking the vehicle in the former case and driving along the track in the latter. In Humanoid, we considered the trajectory of the height of the robot, which determines whether the robot falls or not. Since trajectories can have different lengths, we extended them with zero-padding to the maximum observed length. For both input and output diversity we used the \textit{k-means} clustering algorithm~\cite{kmeans} and we determined the optimal number of clusters $\symks$ by performing \textit{silhouette analysis}~\cite{silhouette}, optimizing the balance between density and separation of the clusters. In our experiments we varied the number of clusters $\symk$ between two and the number of inputs (i.e., environment configurations or trajectories) to be clustered, and computed the silhouette score for each candidate. We increased $\symks$ to a higher value only if the new silhouette was at least 20\% greater than the best silhouette score found so far, in order to filter out random fluctuations of the silhouette score.

After applying clustering, we computed two diversity metrics for each failure search approach, namely \textit{coverage} and \textit{entropy}. Given a failure search approach $\symAg$ and the optimal number of clusters $\symks$,  the coverage of the clusters for the approach $\symAg$ is defined as: 

\begin{equation}
	\symC_{\symAg} = \frac{\sum_{i=1}^{\symks} \symgamma_{\symAg}(i)}{\symks}
\end{equation} 
\noindent
where the function $\symgamma_{\symAg}(i): \mathbb{Z}^+ \rightarrow 0|1$ determines whether a certain cluster labeled by $i$ is \textit{covered} by the failure search approach $\symAg$, i.e., whether at least one failure generated by the approach $\symAg$ belongs to the cluster with label $i$. 

The second metric, \textit{entropy}, measures how uniformly the different failures are distributed across the clusters. Given the number of failures triggered by the failure search approach $\symAg$ in the $i$-th cluster, $\symF_{\symAg}(i)$, the probability of finding a failure generated by the approach $\symAg$ in cluster $i$ is given by: 

\begin{equation}
	\symeta_i^\symAg = \frac{\symF_{\symAg}(i)}{\sum_i \symF_{\symAg}(i)}
\end{equation} 
\noindent
and entropy is defined as: 

\begin{equation}
	\symH_{\symAg} = \sum_{i= 1}^{\symks} \symeta_i^\symAg \cdot \log_2(\symeta_i^\symAg)
\end{equation}

In particular, entropy is zero when all failures are concentrated in one cluster, while it is maximum and equal to $\log_2(\symks)$ when failures are distributed uniformly across all the clusters. Hence, we can use the following formula as a normalized measure of entropy ranging between 0 and 1:

\begin{equation}
	\overline{\symH}_{\symAg} = \frac{\symH_{\symAg}}{\log_2(\symks)}
\end{equation}

We ran clustering ten times to account for randomness and took the average of coverage and entropy across the ten runs. Then, we compared statistically the coverage and entropy averages for each failure search approach across the respective ten runs by computing the Mann-Whitney U Test and the Vargha-Delaney effect size.

To evaluate the impact of algorithm, seed and mutation strategies on \tool (RQ\textsubscript{3}), we used the same metrics: number of failures and input/output diversity, the latter quantified with coverage and entropy.

\subsection{Results}

\begin{table*}[ht]
	\setlength{\tabcolsep}{2pt}
	\renewcommand{\arraystretch}{1.4}
	\centering
	\caption{Number of failures and input/output diversity measured by cluster coverage or entropy. Values represent the average among ten runs. Bold faced values indicate a statistically significant difference between \tool  and sampling. Underlined values indicate a large effect size.}
	\begin{tabular}{l@{\hskip -3em}rccccc@{\hskip 1.5em}ccccc@{\hskip 1.5em}ccccc}
	
		\toprule
		
		\multicolumn{2}{c}{}  
		& \multicolumn{5}{c}{\textsc{Parking}} 
		& \multicolumn{5}{c}{\textsc{Humanoid}} 
		& \multicolumn{5}{c}{\textsc{Driving}} \\
		
		\cmidrule(r){3-7} 
		\cmidrule(r){8-12} 
		\cmidrule(r){13-17}
				
		&  
		& \multicolumn{5}{c}{Diversity} 
		& \multicolumn{5}{c}{Diversity} 
		& \multicolumn{5}{c}{Diversity} \\
		
		\cmidrule(r){4-7}
		\cmidrule(r){9-12}
		\cmidrule(r){14-17}

		&  &  & 
		\multicolumn{2}{c}{Input} 
		& \multicolumn{2}{c}{Output} 
		&  
		& \multicolumn{2}{c}{Input}
		& \multicolumn{2}{c}{Output} 
		&  
		& \multicolumn{2}{c}{Input} 
		& \multicolumn{2}{c}{Output} \\
		
		\cmidrule(r){4-5}
		\cmidrule(r){6-7}
		\cmidrule(r){9-10}
		\cmidrule(r){11-12}
		\cmidrule(r){14-15}
		\cmidrule(r){16-17}
		
		& 
		& \rot{\# Failures}
		& \rot{Coverage (\%)}
		& \rot{Entropy (\%)}
		& \rot{Coverage (\%)}
		& \rot{Entropy (\%)}
		& \rot{\# Failures}
		& \rot{Coverage (\%)}
		& \rot{Entropy (\%)}
		& \rot{Coverage (\%)}
		& \rot{Entropy (\%)}
		& \rot{\# Failures}
		& \rot{Coverage (\%)}
		& \rot{Entropy (\%)}
		& \rot{Coverage (\%)}
		& \rot{Entropy (\%)} \\
		
		\midrule
		
		\textbf{Baselines} & \multicolumn{1}{l}{} & \multicolumn{1}{l}{} & \multicolumn{1}{l}{} & \multicolumn{1}{l}{} & \multicolumn{1}{l}{} & \multicolumn{1}{l}{} & \multicolumn{1}{l}{} & \multicolumn{1}{l}{} & \multicolumn{1}{l}{} & \multicolumn{1}{l}{} & \multicolumn{1}{l}{} & \multicolumn{1}{l}{} & \multicolumn{1}{l}{} & \multicolumn{1}{l}{} & \multicolumn{1}{l}{} & \multicolumn{1}{l}{} \\
		
		& random
		& 1 & 55.00 & 18.37 & 31.94 & 12.13 & 1 & 16.08 & 0.00 & 30.00 & 0.00 & 1 & 10.00 & 0.00 & 10.00 & 0.00 \\
		
		\multicolumn{1}{l}{} 
		& sampling
		& 13 & 50.00 & 0.00 & 41.10 & 22.91 & 1 & 14.72 & 12.62 & 25.00 & 0.00 & 5 & 80.00 & 54.97 & 75.00 & 37.50 \\
		
		\textbf{\tool} & \multicolumn{1}{l}{} & \multicolumn{1}{l}{} & \multicolumn{1}{l}{} & \multicolumn{1}{l}{} & \multicolumn{1}{l}{} & \multicolumn{1}{l}{} & \multicolumn{1}{l}{} & \multicolumn{1}{l}{} & \multicolumn{1}{l}{} & \multicolumn{1}{l}{} & \multicolumn{1}{l}{} & \multicolumn{1}{l}{} & \multicolumn{1}{l}{} & \multicolumn{1}{l}{} & \multicolumn{1}{l}{} & \multicolumn{1}{l}{} \\
		
		& hc\textsubscript{rnd}  
		& 4 & {\ul \textbf{75.00}} & {\ul \textbf{41.54}} & 44.72 & 26.90 & 1 & 29.57 & 19.03 & 40.00 & 0.00 & 3 & 65.00 & 23.80 & 56.00 & 10.11 \\
		
		& hc\textsubscript{fail}  
		& 5 & {\ul \textbf{85.00}} & {\ul \textbf{64.87}} & 45.45 & 33.95 & 2 & 27.37 & 18.00 & \textbf{50.00} & 16.23 & 4 & 90.00 & 73.61 & \textbf{95.00} & 77.20 \\
		
		& hc\textsubscript{sal+rnd}  
		& 6 & {\ul \textbf{90.00}} & {\ul \textbf{57.37}} & 48.01 & 31.59 & {\ul \textbf{4}} & {\ul \textbf{59.73}} & {\ul \textbf{61.32}} & {\ul \textbf{95.00}} & {\ul \textbf{77.75}} & 2 & 60.00 & 28.37 & 60.00 & 26.48 \\
		
		& hc\textsubscript{sal+fail}  
		& 13 & {\ul \textbf{100.00}} & {\ul \textbf{93.23}} & 66.03 & {\ul \textbf{58.91}} & {\ul \textbf{3}} & {\ul \textbf{56.98}} & {\ul \textbf{55.66}} & {\ul \textbf{80.00}} & {\ul \textbf{53.70}} & {\ul \textbf{11}} & \textbf{100.00} & 83.80 & {\ul \textbf{100.00}} & {\ul \textbf{86.69}} \\
		
		& ga\textsubscript{rnd}  
		& 6 & 55.00 & 5.44 & 43.39 & 25.68 & 2 & 35.95 & 31.21 & \textbf{55.00} & \textbf{29.18} & 6 & 90.00 & 72.00 & 80.00 & 52.86 \\
		
		& ga\textsubscript{fail}  
		& 11 & 50.00 & 0.00 & 46.55 & 29.64 & 2 & 24.99 & 12.90 & {\ul \textbf{63.50}} & {\ul \textbf{43.48}} & {\ul \textbf{22}} & 90.00 & 85.63 & 70.00 & 19.16 \\
		
		& ga\textsubscript{sal+rnd}  
		& 8 & 60.00 & 10.31 & 47.73 & 29.77 & 1 & 23.87 & 17.45 & 40.00 & 19.52 & 5 & 83.00 & 59.70 & 60.00 & 17.22 \\
		
		& ga\textsubscript{sal+fail}  
		& {\ul \textbf{27}} & 55.00 & 2.01 & 51.95 & 29.22 & {\ul \textbf{4}} & {\ul \textbf{38.11}} & 24.99 & {\ul \textbf{75.00}} & {\ul \textbf{44.04}} & {\ul \textbf{42}} & \textbf{100.00} & {\ul \textbf{97.61}} & 70.00 & 12.73 \\

		\bottomrule
		
	\end{tabular}
	\label{table:empirical-evaluation:rq1}
\end{table*}

\head{Effectiveness (RQ\textsubscript{1})} Column \textit{\# Failures} in \autoref{table:empirical-evaluation:rq1} shows the average number of failure environment configurations triggered by each failure search approach out of ten runs for each case study. Considering \tool in all its settings, such number is between 4 and 27 in the Parking environment, between 1 and 4 in the Humanoid environment and between 2 and 42 in the Driving environment.

\begin{tcolorbox}[colback=gray!15!white,colframe=black]
	\textbf{RQ\textsubscript{1}}: Overall, \tool successfully challenged the DRL agent under test in all case studies, by generating a significant number of failure environment configurations. On average, the sampling approach generated 6 failures while \tool generated from 3 to 24 failures.
\end{tcolorbox}

\head{Comparison (RQ\textsubscript{2})} \autoref{table:empirical-evaluation:rq1} shows the results for the first setting: bold values indicate a statistically significant difference (at level $\symalpha = 0.05$) between \tool  and sampling; values are underlined when the effect size is large. In particular, in Parking the best approach is ga\textsubscript{sal+fail} that is able to generate more failures than other approaches (i.e., 27 on average) and the difference w.r.t. the sampling approach (i.e., 13 failures on average) is statistically significant with a large effect size. In Humanoid, several settings of \tool are significantly better than sampling (which exposes only 1 failure on average). In Driving, ga\textsubscript{sal+fail} and hc\textsubscript{sal+fail} are the best approaches with, on average, 42 and 11 failures respectively. Their difference w.r.t. the sampling approach (which exposes 5 failures on average) is statistically significant, and the effect size is large.  

Regarding the comparison with the random approach, in all case studies, classifier-based approaches (i.e., sampling and \tool) found significantly more failures with a large effect size. Hence, our empirical results show that the failures experienced during training are indeed related to and informative of the failures of the DRL agent under test.

The macro-columns \textit{Diversity} in \autoref{table:empirical-evaluation:rq1} show the average out of ten runs of the \textit{Coverage} and \textit{Entropy} metrics regarding input and output diversity for each failure search approach. Although in Parking the ga\textsubscript{sal+fail} approach generates more failures than sampling, the two approaches are comparable in terms of input and output diversity (both considering coverage and entropy). On the other hand, the hc\textsubscript{sal+fail} approach is able to generate inputs that are both significantly different from those of sampling (i.e., higher coverage) and better distributed among clusters (i.e., higher entropy). Output coverage and entropy values are higher for hc\textsubscript{sal+fail} than those of sampling (66.03\% and 58.91\% vs 41.10\% and 22.91\% respectively), with output coverage being comparable and output entropy being statistically better with a large effect size. This means that, although hc\textsubscript{sal+fail} and sampling generated the same number of failures (i.e., 13 on average), the failures produced by hc\textsubscript{sal+fail} exercise more diverse behaviours of the DRL agent under test.

For what concerns Humanoid, the two best approaches in terms of input and output diversity are settings of \tool (hc\textsubscript{sal+rnd} and hc\textsubscript{sal+fail}) both considering coverage and entropy (the ga\textsubscript{sal+fail} setting is comparable to them). All of them are significantly better than sampling with a large effect size, except for ga\textsubscript{sal+fail}, whose input entropy (24.99\%) is not significantly different from that of sampling (12.62\%). In Driving, the best approaches in terms of input and output diversity are hc\textsubscript{sal+fail} and ga\textsubscript{sal+fail}, the former significantly better than sampling on input coverage and output diversity and the latter on input diversity.

When comparing \tool with the random approach in terms of diversity, most of the \tool settings, especially those with the saliency-based mutations, are significantly better than random with a large effect size in all case studies.

We identify two reasons for the higher diversity achieved by \tool: (1)~the search-based approach of \tool implicitly favors diversity by locally exploring the environment configuration space until failures are discovered. In this way, it is more likely for \tool to obtain failure configurations starting from diverse seeds. This is more difficult when failure configurations are to be randomly generated from scratch (e.g., by the sampling approach). (2)~The failure environment configurations generated at training time affect different versions of the agent under test, which is evolving during its training process. Hence, these configurations tend to be quite diverse. \tool effectively uses such seed configurations by retaining their diversity while at the same time increasing their failure exposure.

\begin{tcolorbox}[colback=gray!15!white,colframe=black]
	\textbf{RQ\textsubscript{2}}: Overall, both considering the number of failures triggered and their input and output diversity, hc\textsubscript{sal+fail} is the best setting of \tool, generating 50\% more failures than the sampling approach as well as  failures being 78\% more input-diverse and 74\% more output-diverse.
\end{tcolorbox}

\begin{table*}[ht]
	\setlength{\tabcolsep}{2pt}
	\renewcommand{\arraystretch}{1.2}
	\centering
	\caption{Impact of the seed strategy. In each row, the three case studies (Parking / Humanoid / Driving)  are separated by a forward slash ``/'' symbol. An ``F'' symbol indicates a statistically significant difference in favor of the approach with the failure seed; a dash ``-' symbol indicates that the two approaches are indistinguishable and an ``R'' symbol (missing in the table) would indicate a statistically significant difference in favor of the approach with the random seed. As for the comparisons in the first two research questions, we compute statistical significance using the Mann-Whitney U Test.}
	\begin{tabular}{r@{\hskip 1em}c@{\hskip 1em}c@{\hskip 1em}c@{\hskip 1em}c@{\hskip 1em}c}
	
		\toprule
		
		& \multicolumn{5}{c}{\textsc{Parking} / \textsc{Humanoid} / \textsc{Driving}} \\
		
		\cmidrule(r){2-6}
		
		&  
		& \multicolumn{4}{c}{Diversity} \\
		
		\cmidrule(r){3-6}
		
		&  
		& \multicolumn{2}{c}{Input} 
		& \multicolumn{2}{c}{Output} \\
		
		\cmidrule(r){3-4}
		\cmidrule(r){5-6}
		
		& \rot{\# Failures} 
		& \rot{Coverage \,\,} 
		& \rot{Entropy} 
		& \rot{Coverage \,\,} 
		& \rot{Entropy} \\
		
		\midrule
		
		hc\textsubscript{rnd} \textit{vs} hc\textsubscript{fail}  & - / - / F & - / - / F & - / - / F & - / - / F & - / - / F \\
		
		ga\textsubscript{rnd} \textit{vs} ga\textsubscript{fail}  & F / - / F & - / - / - & - / - / - & - / - / - & - / - / - \\
		
		hc\textsubscript{sal+rnd} \textit{vs} hc\textsubscript{sal+fail}  & F  / - / F & - / - / F & F / - / F & - / - / F & F / - / F \\
		
		ga\textsubscript{sal+rnd} \textit{vs} ga\textsubscript{sal+fail}  & F / F / F & - / - / F & - / - / F & - / F / - & - / - / - \\
		
		\midrule
		
		\textbf{rnd} \textit{vs} \textbf{fail} & 0 \textit{vs} \textbf{8} & 0 \textit{vs} \textbf{3} & 0 \textit{vs} \textbf{4} & 0 \textit{vs} \textbf{3} & 0 \textit{vs} \textbf{3} \\
		
		\bottomrule
		
	\end{tabular}
	\label{table:empirical-evaluation:rq2}
\end{table*}

\head{Hyperparameters (RQ\textsubscript{3})} 
On average and considering all case studies, hc\textsubscript{sal+fail} generates failures that are the most diverse both in terms of input and output diversity. Specifically, hc\textsubscript{sal+fail} has the best input coverage (85.66\% vs 69.91\% of the second best hc\textsubscript{sal+rnd}), the best input entropy (77.56\% vs 52.16\% of the second best hc\textsubscript{fail}), the best output coverage (82.01\% vs 67.67\% of the second best hc\textsubscript{sal+rnd}) and the best output entropy (66.43\% vs 45.27\% of the second best hc\textsubscript{sal+rnd}). 

On the other hand, ga\textsubscript{sal+fail} is the setting of \tool that generates the highest number of failures (i.e., 24 on average). However, such failures cover fewer clusters and have a lower entropy than the failures generated by hc\textsubscript{sal+fail}. %
Therefore, considering the number of failures and their diversity, hc\textsubscript{sal+fail} is the preferable \tool setting.

Across all case studies, the settings of \tool that use the saliency-based mutations are more effective than their counterparts, i.e., the difference is always statistically significant with a large effect size, except for Humanoid, where ga\textsubscript{sal+fail} is comparable to ga\textsubscript{rnd}. Overall, this shows that the guidance offered by the saliency-based mutation operator is effective at finding failure environment configurations.

In \autoref{table:empirical-evaluation:rq2} we report a further comparison between the different settings of \tool, focused on the impact of the seed strategy (random, \textit{rnd}, vs failure, \textit{fail}). 
In each cell of the table the three case studies are separated by a forward slash ``/'' symbol.  The symbol ``F'' (respectively ``R'') indicates that failure seeding (respectively random seeding) is statistically better than random (respectively failure) seeding. A dash symbol indicates no statistically significant difference. For instance, for the raw \textit{hc\textsubscript{rnd} vs hc\textsubscript{fail} and the macro-column \textit{\# Failures}, the hc\textsubscript{rnd} and the hc\textsubscript{fail} settings are equivalent in the first two case studies, i.e., Parking and Humanoid (hence the two ``-'' symbols separated by the ``/'' symbol), while the hc\textsubscript{fail} setting is statistically better than hc\textsubscript{rnd} in the last case study, i.e., Driving (hence the last symbol is ``F'').}
In terms of number of failures we can see that, in most cases (i.e., 8/12) the settings with the failure seeds are significantly better than their random seeds counterparts. In particular, the ga\textsubscript{sal+fail} setting is significantly better than the ga\textsubscript{sal+rnd} setting in all case studies (while, for instance, the hc\textsubscript{sal+fail} is significantly better than hc\textsubscript{sal+rnd} in Parking and Driving but not in Humanoid). From the point of view of diversity, the settings of \tool with the failure seeds are mostly comparable to the settings with the random seeds, except for Driving, where the failure seeds are critical to generate more diverse failures. Random seeding is never a better choice (indeed the ``R'' symbol is completely missing in the table), across all settings of \tool and across all three case studies.

\begin{tcolorbox}[colback=gray!15!white,colframe=black]
	\textbf{RQ\textsubscript{3}}: Overall, the settings of \tool with the saliency mutation strategy and failure seeds are either comparable or significantly better than their random counterparts. In particular, failure seeds are significantly better than random seeds in 21 comparisons out of 60. Between hill climbing and genetic algorithm, the former is preferable because it generates more diverse failure scenarios (i.e., on average hill climbing failures are 60\% more input-diverse and 80\% more output-diverse than genetic algorithm failures). The latter might be considered when the number of exposed failures is important, regardless of their diversity.
\end{tcolorbox}

\begin{table*}[ht]
	\setlength{\tabcolsep}{2pt}
	\renewcommand{\arraystretch}{1.4}
	\centering
	\caption{Number of failures and input/output diversity measured by cluster coverage or entropy. Values represent the average among ten runs. Bold faced values indicate a statistically significant difference between \tool and the best of the two baselines. Underlined values indicate a large effect size. We indicate with a bar the setting of \tool using the fitness-based surrogate model, i.e., $\overline{\text{hc}}$\textsubscript{sal+rnd}}.
	\begin{tabular}{l@{\hskip -3em}rccccc@{\hskip 1.5em}ccccc@{\hskip 1.5em}ccccc}
	
		\toprule
		
		\multicolumn{2}{c}{}  
		& \multicolumn{5}{c}{\textsc{Parking}} 
		& \multicolumn{5}{c}{\textsc{Humanoid}} 
		& \multicolumn{5}{c}{\textsc{Driving}} \\
		
		\cmidrule(r){3-7} 
		\cmidrule(r){8-12} 
		\cmidrule(r){13-17}
				
		&  
		& \multicolumn{5}{c}{Diversity} 
		& \multicolumn{5}{c}{Diversity} 
		& \multicolumn{5}{c}{Diversity} \\
		
		\cmidrule(r){4-7}
		\cmidrule(r){9-12}
		\cmidrule(r){14-17}
		
		&  &  & 
		\multicolumn{2}{c}{Input} 
		& \multicolumn{2}{c}{Output} 
		&  
		& \multicolumn{2}{c}{Input}
		& \multicolumn{2}{c}{Output} 
		&  
		& \multicolumn{2}{c}{Input} 
		& \multicolumn{2}{c}{Output} \\
		
		\cmidrule(r){4-5}
		\cmidrule(r){6-7}
		\cmidrule(r){9-10}
		\cmidrule(r){11-12}
		\cmidrule(r){14-15}
		\cmidrule(r){16-17}
		
		& 
		& \rot{\# Failures}
		& \rot{Coverage (\%)}
		& \rot{Entropy (\%)}
		& \rot{Coverage (\%)}
		& \rot{Entropy (\%)}
		& \rot{\# Failures}
		& \rot{Coverage (\%)}
		& \rot{Entropy (\%)}
		& \rot{Coverage (\%)}
		& \rot{Entropy (\%)}
		& \rot{\# Failures}
		& \rot{Coverage (\%)}
		& \rot{Entropy (\%)}
		& \rot{Coverage (\%)}
		& \rot{Entropy (\%)} \\
		
		\midrule
		
		\textbf{Baselines} & \multicolumn{1}{l}{} & \multicolumn{1}{l}{} & \multicolumn{1}{l}{} & \multicolumn{1}{l}{} & \multicolumn{1}{l}{} & \multicolumn{1}{l}{} & \multicolumn{1}{l}{} & \multicolumn{1}{l}{} & \multicolumn{1}{l}{} & \multicolumn{1}{l}{} & \multicolumn{1}{l}{} & \multicolumn{1}{l}{} & \multicolumn{1}{l}{} & \multicolumn{1}{l}{} & \multicolumn{1}{l}{} & \multicolumn{1}{l}{} \\
		
		& random
		& 1 & 46.67 & 12.10 & 55.00 & 19.18 & 1 & 30.83 & 0.00 & 37.50 & 5.00 & 0 & 23.72 & 0.00 & 20.00 & 0.00 \\
		
		\multicolumn{1}{l}{} 
		& train. fail.
		& 1 & 45.00 & 0.00 & 50.00 & 0.00 & 1 & 25.83 & 0.00 & 35.00 & 10.00 & 2 & 33.34 & 9.70 & 60.00 & 20.00 \\
		
		\textbf{\tool} & \multicolumn{1}{l}{} & \multicolumn{1}{l}{} & \multicolumn{1}{l}{} & \multicolumn{1}{l}{} & \multicolumn{1}{l}{} & \multicolumn{1}{l}{} & \multicolumn{1}{l}{} & \multicolumn{1}{l}{} & \multicolumn{1}{l}{} & \multicolumn{1}{l}{} & \multicolumn{1}{l}{} & \multicolumn{1}{l}{} & \multicolumn{1}{l}{} & \multicolumn{1}{l}{} & \multicolumn{1}{l}{} & \multicolumn{1}{l}{} \\
		
		& hc\textsubscript{sal+rnd}  
		& {\ul \textbf{3}} & {\ul \textbf{71.67}} & {\ul \textbf{57.22}} & \textbf{80.00} & \textbf{62.42} & {\ul \textbf{3}} & 43.42 & 7.29 & {\ul \textbf{75.00}} & \textbf{48.08} & 3 & 57.97 & {\ul \textbf{48.61}} & 80.00 & 56.23 \\
		
		& $\overline{\text{hc}}$\textsubscript{sal+rnd}  
		& {\ul \textbf{3}} & {\ul \textbf{80.00}} & {\ul \textbf{62.06}} & \textbf{80.00} & 52.88 & {\ul \textbf{12}} & {\ul \textbf{81.92}} & {\ul \textbf{57.35}} & {\ul \textbf{100.00}} & {\ul \textbf{86.02}} & {\ul \textbf{10}} & {\ul \textbf{91.16}} & {\ul \textbf{88.10}} & {\ul \textbf{100.00}} & 80.48 \\
		
		\bottomrule
		
	\end{tabular}
	\label{table:empirical-evaluation:rq4}
\end{table*}

\head{Ineffective Failure Seeds (RQ\textsubscript{4})} RQ\textsubscript{3} shows that the saliency mutation operator is effective at guiding the search towards failure-inducing configurations and that hill climbing generates more diverse failures w.r.t. genetic algorithm. To test the effectiveness of \tool in the presence of ineffective failure seeds, we consider its best configuration, i.e., hill climbing with the saliency mutation operator and random seeds. We chose to use random seeds to test whether the surrogate model is able to guide the search without relying on ineffective failure seeds. In \autoref{table:empirical-evaluation:rq4}, we indicate with hc\textsubscript{sal+rnd} the version of \tool using the classifier, while we indicate with  $\overline{\text{hc}}$\textsubscript{sal+rnd} the version of \tool that uses the regressor-based surrogate model.

\autoref{table:empirical-evaluation:rq4} shows that replaying training failures (row \textit{train. fail.}) is ineffective at testing time, being indistinguishable from random in Parking and Humanoid. In Driving,  training failure replay generates significantly more failures than random (i.e., on average 2 vs 0 respectively), although regarding diversity, only the output coverage of training failures is statistically different than the output coverage of failures generated by random.

Regarding \tool, the version with the classifier-based surrogate model (i.e., hc\textsubscript{sal+rnd}) generates significantly more failures than random and training failure replay in Parking and Humanoid. Such failures are also significantly more diverse than the two baselines in Parking and Humanoid, although in the latter there is statistical significance only for output coverage. Regarding Driving, hc\textsubscript{sal+rnd} failures have a significantly higher input entropy, while for output coverage there is no statistical significance, although the trend is in favor of hc\textsubscript{sal+rnd} (i.e., 80 vs 60 for output coverage and 56 vs 20 for output entropy).

The version of \tool with the regressor-based surrogate model (i.e., $\overline{\text{hc}}$\textsubscript{sal+rnd}) is equally effective at finding failures as the classifier-based one in Parking. In Humanoid and Driving, it shows statistical significance of the differences w.r.t. the two baselines, while the classifier-based one does not (e.g., on input diversity in Humanoid; number of failures, input and output coverage in Driving).

\begin{tcolorbox}[colback=gray!15!white,colframe=black]
	\textbf{RQ\textsubscript{4}}: Overall, when failure seeds are ineffective, \tool is able to trigger more failures of the DRL agent under test than replaying the training configurations at testing time in all three case studies (i.e., on average 3 vs 1 in Parking, 12 vs 1 in Humanoid and 10 vs 2 in Driving), with the differences being statistically significant with a large effect size.
\end{tcolorbox}
	\section{Discussion} \label{sec:discussion}

\subsection{Solvability of the Failures}
For each approach, we resumed training of the three DRL agents under test by feeding all the failure environment configurations found by  each approach in every run of the environment generation process. In every case study and for each approach, we found that the given DRL agent could learn how to successfully terminate the respective episodes by performing some additional training. This shows that the failure environment configurations generated by all approaches, and indeed by \tool, are \textit{solvable} by the DRL agents under test. 

This also indicates that, although a generated environment configuration can be challenging for the DRL agent under test, there exists a sequence of actions that let the DRL agent solve the task successfully. For instance, after additional training, the DRL agent in the Parking environment is always able to find trajectories for the vehicle to reach the target spot, which is instead missed by the original DRL agent under test. Similarly, in the Driving environment, the generated failure environment configurations do not induce track shapes that are beyond the mechanical capabilities of the vehicle. Regarding Humanoid, the initial positions of the joints as well as their velocities, resulting from the generated failure environment configurations, do not prevent the DRL agent to control and regain the balance of the robot.

Solvability of the failure environment configurations generated by \tool shows that such environment configurations represent real weaknesses of the DRL agent under test, which could realistically occur during the operation of the DRL agent in production.

\subsection{Training Failures}

\subsubsection{Qualitative Analysis}
A simple way to test a DRL agent is to replay the training failures except for the earliest ones (e.g., by filtering out the first 30\% of the failure environment configurations). However, this has two major downsides: (1)~the DRL agent under test may have adapted to failure environment configurations in which weaker versions of itself failed, despite the exclusion of early failures; (2)~such an approach would not be generative and by design it can only replay existing failure environment configurations. Generative approaches like hc\textsubscript{sal+fail} produce diverse and potentially unlimited challenging inputs exposing the limitations of the DRL agent, and potentially improving it through retraining~\cite{everett-thesis}.

\begin{figure*}[t]
\centering

\includegraphics[trim=1cm 20cm 15cm 1cm, clip=true, scale=0.25]
{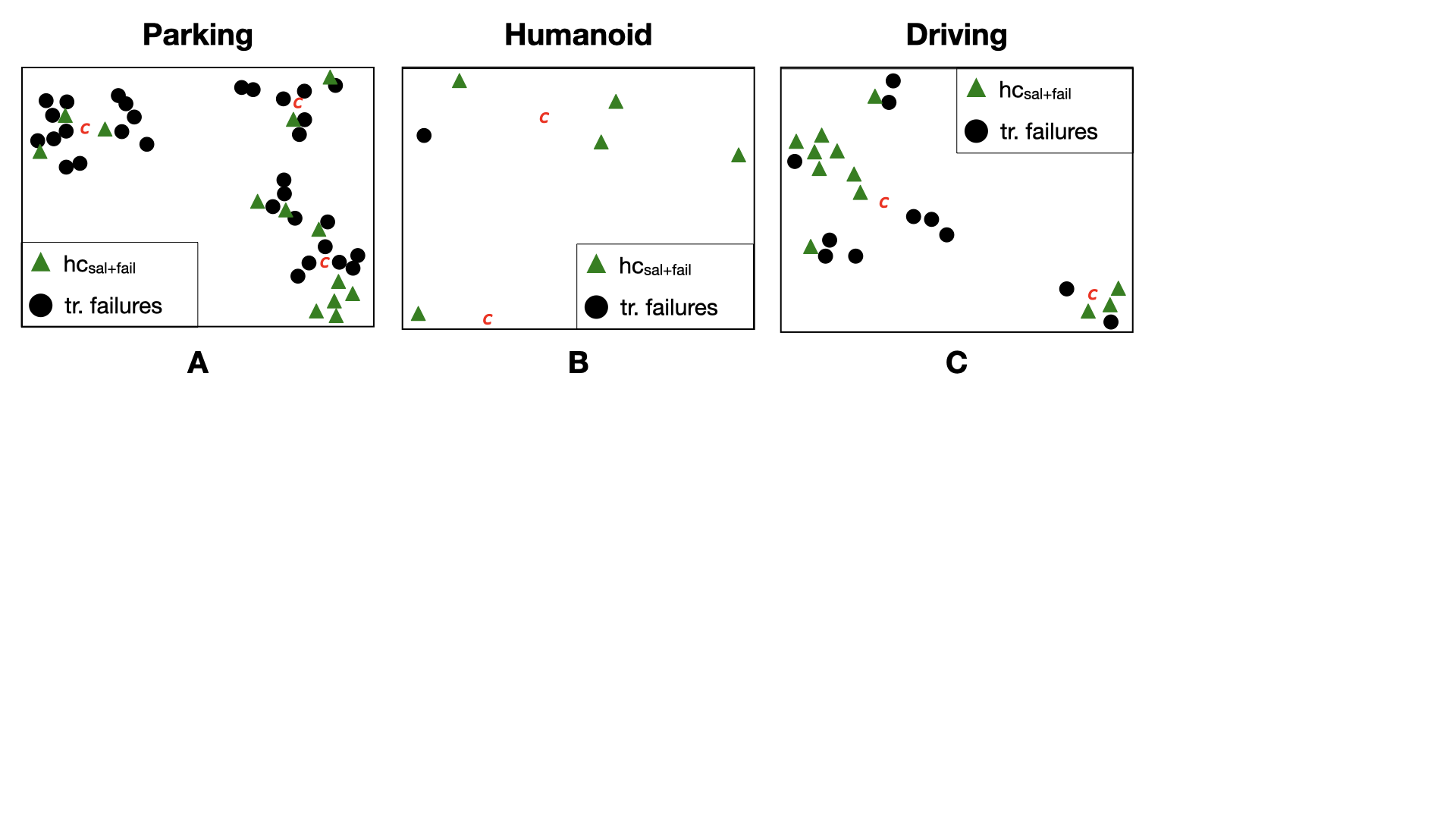}

\caption{t-SNE projection in 2D of trajectories associated with hc\textsubscript{sal+fail} (\protect\tikz \fill[color=ao] (0,0) -- (0.2,0) -- (0.1,0.2) -- cycle;) and training failures (\protect\tikz \protect\draw[fill=black] (0,0) circle (0.1cm);) configurations. The \textcolor{red}{\textbf{\textit{C}}} symbol indicates the centroid of a cluster. The leftmost plot (\textbf{A}) shows trajectories for the Parking environment, the center plot (\textbf{B}) for the Humanoid environment, and the rightmost plot (\textbf{C}) for the Driving environment.} 
\label{fig:empirical-evaluation:clustering-abstract} 
\end{figure*}

For the sake of completeness, we replayed at testing time the failure environment configurations that happened during the DRL training process, and we compared them with the failures generated by hc\textsubscript{sal+fail}. In particular, we clustered the DRL agents trajectories associated with those failures, following the same process described in \autoref{sec:evaluation:metrics}. \autoref{fig:empirical-evaluation:clustering-abstract} shows a 2D t-SNE~\cite{tsne} projection of the failure trajectories of the three DRL agents under all failure environment configurations in each case study. Cluster centroids are indicated with the letter \textcolor{red}{\textbf{\textit{C}}}. In Humanoid (\autoref{fig:empirical-evaluation:clustering-abstract}.\textbf{B}) the trajectories associated with the failure environment configurations discovered by \tool, cover more clusters than the trajectories associated with \textit{training} failure environment configurations. In Parking (\autoref{fig:empirical-evaluation:clustering-abstract}.\textbf{A}) and Driving (\autoref{fig:empirical-evaluation:clustering-abstract}.\textbf{C}) the two classes of trajectories are complementary since they cover the same clusters but with a different intra-cluster distribution, showing that the generative approach of \tool can explore new clusters or new regions within a cluster.

\begin{table*}[ht]
	\setlength{\tabcolsep}{1.5pt}
	\renewcommand{\arraystretch}{1.4}
	\footnotesize
	\centering
	\caption{Comparison between training failures (i.e., train. fail.) and failures found by the best \tool settings (i.e., hc\textsubscript{sal+fail} and ga\textsubscript{sal+fail}). The comparison is in terms of number of failures, input/output diversity measured by cluster coverage or entropy and Gini purity coefficient. Values represent the average among ten runs. Bold faced values indicate a statistically significant difference. Underlined values indicate a large effect size.}
	\begin{tabular}{l@{\hskip 1em}ccccccc@{\hskip 1em}ccccccc@{\hskip 1em}ccccccc}
	
		\toprule
		
		& \multicolumn{6}{c}{\textsc{Parking}} &  
		& \multicolumn{6}{c}{\textsc{Humanoid}} & 
		& \multicolumn{6}{c}{\textsc{Driving}} &  \\
		
		\cmidrule(r){2-8} 
		\cmidrule(r){9-15} 
		\cmidrule(r){16-22}
				
		&  
		& \multicolumn{6}{c}{Diversity} 
		& 
		& \multicolumn{5}{c}{Diversity} &  
		& 
		& \multicolumn{5}{c}{Diversity} &  \\
		
		\cmidrule(r){3-8}
		\cmidrule(r){10-15}
		\cmidrule(r){17-22}

		& \multicolumn{1}{l}{} 
		& \multicolumn{3}{c}{Input} 
		& \multicolumn{3}{c}{Output} 
		& \multicolumn{1}{l}{} 
		& \multicolumn{3}{c}{Input} 
		& \multicolumn{3}{c}{Output} 
		& \multicolumn{1}{l}{} 
		& \multicolumn{3}{c}{Input} 
		& \multicolumn{3}{c}{Output} \\
		
		\cmidrule(r){3-5}
		\cmidrule(r){6-8}
		\cmidrule(r){10-12}
		\cmidrule(r){13-15}
		\cmidrule(r){17-19}
		\cmidrule(r){20-22}

		& \rot{\# Failures} 
		& \rot{Coverage (\%)} 
		& \rot{Entropy (\%)} 
		& \rot{Gini Purity (\%)} 
		& \rot{Coverage (\%)} 
		& \rot{Entropy (\%)} 
		& \rot{Gini Purity (\%)} 
		& \rot{\# Failures} 
		& \rot{Coverage (\%)} 
		& \rot{Entropy (\%)} 
		& \rot{Gini Purity (\%)} 
		& \rot{Coverage (\%)} 
		& \rot{Entropy (\%)} 
		& \rot{Gini Purity (\%)} 
		& \rot{\# Failures} 
		& \rot{Coverage (\%)} 
		& \rot{Entropy (\%)} 
		& \rot{Gini Purity (\%)} 
		& \rot{Coverage (\%)} 
		& \rot{Entropy (\%)} 
		& \rot{Gini Purity (\%)} \\

		\midrule
		
		train. fail. & {\ul \textbf{31}} & 100.00 & 98.10 & \multirow{2}{*}{60.48} & 95.00 & 40.62 & \multirow{2}{*}{66.93} & 1 & 35.00 & 10.00 & \multirow{2}{*}{83.42} & 35.00 & 10.00 & \multirow{2}{*}{87.40} & 13 & 100.00 & 78.32 & \multirow{2}{*}{53.84} & 100.00 & 74.66 & \multirow{2}{*}{53.23} \\
		
		hc\textsubscript{sal+fail} & 13 & 100.00 & 93.53 &  & 85.00 & 42.24 &  & {\ul \textbf{3}} & {\ul \textbf{100.00}} & {\ul \textbf{66.19}} &  & {\ul \textbf{100.00}} & {\ul \textbf{63.70}} &  & 11 & 100.00 & 81.95 &  & 100.00 & 83.40 &  \\
		& \multicolumn{1}{l}{} & \multicolumn{1}{l}{} & \multicolumn{1}{l}{} & \multicolumn{1}{l}{} & \multicolumn{1}{l}{} & \multicolumn{1}{l}{} & \multicolumn{1}{l}{} & \multicolumn{1}{l}{} & \multicolumn{1}{l}{} & \multicolumn{1}{l}{} & \multicolumn{1}{l}{} & \multicolumn{1}{l}{} & \multicolumn{1}{l}{} & \multicolumn{1}{l}{} & \multicolumn{1}{l}{} & \multicolumn{1}{l}{} & \multicolumn{1}{l}{} & \multicolumn{1}{l}{} & \multicolumn{1}{l}{} & \multicolumn{1}{l}{} & \multicolumn{1}{l}{} \\
		
		train. fail. & {\ul \textbf{31}} & {\ul \textbf{100.00}} & {\ul \textbf{85.32}} & \multirow{2}{*}{81.55} & {\ul \textbf{89.83}} & {\ul \textbf{51.49}} & \multirow{2}{*}{81.54} & 1 & 28.33 & 10.00 & \multirow{2}{*}{89.84} & 30.00 & 10.00 & \multirow{2}{*}{89.41} & 13 & 95.00 & 66.99 & \multirow{2}{*}{72.76} & {\ul \textbf{100.00}} & {\ul \textbf{75.57}} & \multirow{2}{*}{69.63} \\
		
		ga\textsubscript{sal+fail} & 27 & 50.00 & 0.00 &  & 48.84 & 25.17 &  & {\ul \textbf{4}} & {\ul \textbf{91.67}} & {\ul \textbf{54.02}} &  & {\ul \textbf{90.00}} & {\ul \textbf{50.51}} &  & {\ul \textbf{42}} & 100.00 & {\ul \textbf{97.61}} &  & 64.50 & 6.78 & \\
		
		\bottomrule
		
	\end{tabular}
	\label{table:discussion:training-failures}
\end{table*}

\subsubsection{Quantitative Analysis}

We also carried out a quantitative analysis by measuring the number of failures triggered when replaying the failure environment configurations that happened during training and their diversity w.r.t. the failure configurations found by the two best  settings of \tool, i.e., hc\textsubscript{sal+fail} and ga\textsubscript{sal+fail}. \autoref{table:discussion:training-failures} shows the results for each case study.

In Parking, the training failure configurations trigger significantly more failures than both \tool's settings (i.e., 31 vs 13 and 27 respectively). Such failures are equivalent to those found by hc\textsubscript{sal+fail} in terms of both input and output diversity, but significantly more diverse than those found by ga\textsubscript{sal+fail}. In Humanoid, both \tool's settings are better than replaying failure configurations in all dimensions, i.e., in terms of number of failures and their diversity. In Driving, hc\textsubscript{sal+fail} and the training failures are equivalent, while ga\textsubscript{sal+fail} is significantly better in terms of number of failures (i.e., 42 vs 13) and their input entropy (i.e., 97.61 vs 66.99), but significantly worse in terms of output diversity.

Since the comparison between failure replay and \tool is not conclusive on input and output diversity (i.e., none of the two is superior w.r.t. the other in all case studies), we measured the \textit{complementarity} between the two competing approaches. In fact, high diversity could be achieved by the two approaches either with similar or with complementary distribution of the inputs/outputs among the same clusters. To measure such complementarity between training failures and the failures that \tool generates, we use the Gini purity coefficient~\cite{gini-impurity} for each cluster,  defined as follows:

\begin{equation} \label{equation:discussion:gini}
	\symG(i) = \sum_{\symfc = 1}^{\symCbar} p^2_{\symfc}
\end{equation}
\noindent
where $i$ indicates the $i$-th cluster, $\symCbar$ is the number of classes and $p_{\symfc}$ represents the probability of finding a data point of class $\symfc$ in cluster $i$. The number of classes is $\symCbar = 2$, since we are measuring the complementarity of two approaches. \autoref{equation:discussion:gini} gives  the probability of having a failure in a cluster belonging to one specific class. In particular, $\symG(i) = 1$ means that  cluster $i$ is pure, i.e., all the data points in the cluster belong to a single class.

The \textit{Gini Purity} columns in \autoref{table:discussion:training-failures} show the average Gini purity coefficients for each case study. Overall, we can observe that, in both settings of \tool, i.e., hc\textsubscript{sal+fail} and ga\textsubscript{sal+fail}, the majority of the clusters are highly pure (i.e., the Gini purity is always greater than 50\%). In particular, the setting ga\textsubscript{sal+fail} shows a higher degree of complementarity with the training failures than hc\textsubscript{sal+fail} (across all the case studies, the Gini purity coefficients are 80.79\% and 67.71\% respectively).

Hence, we conclude that replayed failures and new failures generated by \tool are highly complementary. Developers should use both during RL testing.

\begin{figure*}[t]
\centering

\includegraphics[trim=0cm 4cm 17cm 0cm, clip=true, scale=0.25]
{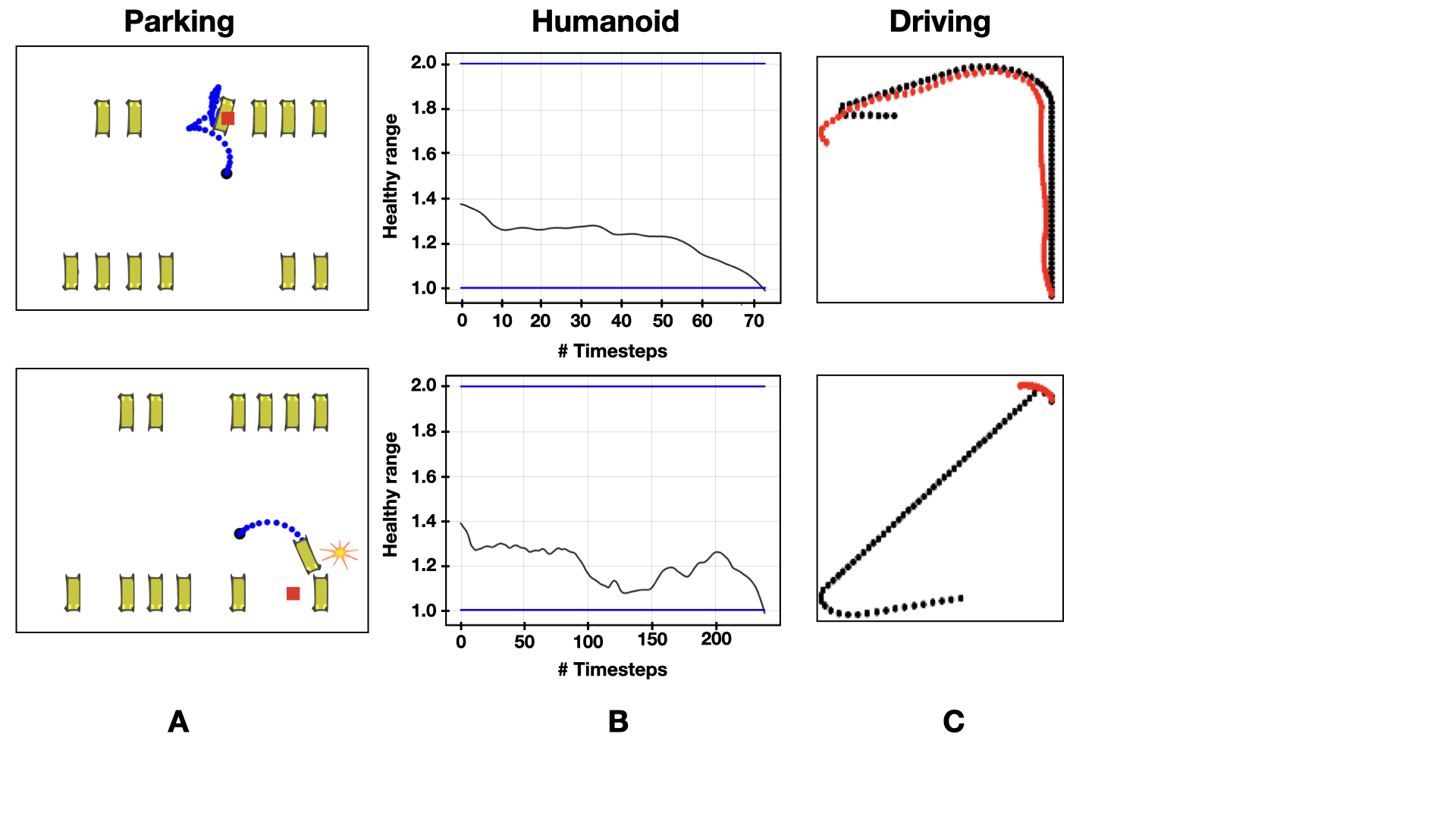}

\caption{Representation of different trajectories associated with hc\textsubscript{sal+fail} failure configurations. The trajectory of the vehicle is shown in blue for the Parking environment (\textbf{A}) and in red for the Driving environment (\textbf{C}). In Humanoid (\textbf{B}), the trajectory of the robot is the latitudinal position over time.} 
\label{fig:empirical-evaluation:clustering-examples} 
\end{figure*}

\subsection{Qualitative Analysis of \tool Failures} \autoref{fig:empirical-evaluation:clustering-examples} shows the failure trajectories associated to failure environment configurations generated by \tool. For all case studies we selected two failure trajectories belonging to different clusters representing two different failure modes. 

In Parking (\autoref{fig:empirical-evaluation:clustering-examples}.\textbf{A}), the first failure mode (at the top) consists of the DRL agent not being able to park the vehicle in the target spot with the right heading given a certain amount of time; the second failure mode (at the bottom) is when the DRL agent does not complete the parking maneuver due to a crash with another vehicle parked beside the target spot. 

In Humanoid (\autoref{fig:empirical-evaluation:clustering-examples}.\textbf{B}), the abdomen latitudinal coordinate of the robot should be among the two horizontal lines shown in \autoref{fig:empirical-evaluation:clustering-examples} called \textit{healthy range}~\cite{humanoid-wiki}, otherwise the robot falls and the episode terminates unsuccessfully. The two failure trajectories associated with \tool failure environment configurations are different: the one at the top shows a trajectory that monotonically goes down in a relatively short amount of time (i.e., just above 70 timesteps). The trajectory at the bottom is more noisy indicating that the DRL agent under test is more uncertain. Indeed, the DRL agent is able to recover the partial fall around the middle of the episode (i.e., around 120 timesteps) while eventually failing after $\approx$ 200 timesteps. 

In Driving (\autoref{fig:empirical-evaluation:clustering-examples}.\textbf{C}), the first failure mode shows the DRL agent failing to drive the last curve of the track, while the trajectory at the bottom represents a track with a difficult left curve at the beginning.

\subsection{Discrete vs Continuous Configurations} In the Parking environment, the number of failures found by the sampling approach are equivalent to the failures found by the best \tool setting, i.e., hc\textsubscript{sal+fail} (i.e., 13 on average), although the failures generated by hc\textsubscript{sal+fail} have a significantly higher input diversity and output entropy. Parking environment configurations are composed of several discrete and only a few continuous parameters, and they are subject to few constraints. As a consequence, sampling Parking environment configurations at random is efficient, as a large number of candidate environment configurations can be generated within the given search budget, among which it is more likely to find environment configurations where the failure prediction is high.

Also in the Humanoid environment there are few validity constraints, hence it is efficient to sample environment configurations at random. However, the space of environment configurations is larger than in Parking (almost twice as much, i.e., 47 vs 24 parameters) and all parameters are in the continuous domain. As a consequence, finding challenging environment configurations by sampling at random is not effective. Indeed, our results show that the sampling approach generates, on average, 1 failure, while hc\textsubscript{sal+fail} generates 3 failures (a 200\% increase), with a significantly higher input and output diversity. This shows that \tool works much better than random sampling when environment configurations have continuous parameters.

The Driving environment offers yet another perspective. Indeed, similarly to Parking, Driving environment configurations are mostly composed of discrete parameters but, on the other hand, such environment configurations are subject to complex constraints. Therefore, sampling Driving environment configurations at random is not efficient, contrary to modifying existing environment configurations. Indeed, the sampling approach generates, on average, 5 failures, while hc\textsubscript{sal+fail} generates 11 failures (i.e., a 120\% increase), with a significantly higher input and output diversity. Although, the Driving environment configurations have mostly discrete parameters our search-based approach is more effective than sampling, since it is able to use the search budget more efficiently by modifying existing environment configurations which satisfy the complex constraints that hold in this environment.

\subsection{Ablation Study on the Driving Environment}

\begin{table*}[ht]
	\setlength{\tabcolsep}{2pt}
	\renewcommand{\arraystretch}{1.1}
	\centering
	\caption{Comparison between failures found by \tool settings on the Driving environment with and without enabling the road constraints when generating new configurations}
	\begin{tabular}{lr@{\hskip 1em}r@{\hskip 1em}c@{\hskip -1em}c}
	
		\toprule
		
		\multicolumn{1}{l}{} &  
		& \multicolumn{2}{c}{\textsc{Driving}} \\
		
		\cmidrule(r){3-4}

		\multicolumn{2}{l}{} 
		& \rot{\textit{w/} Constr.} 
		& \rot{\textit{w/o} Constr.} \\
		
		\multicolumn{1}{l}{} &  
		& \multicolumn{2}{c}{\# Failures} \\
		
		\midrule
		
		\textbf{\tool} &  &  &  \\
		
		& sampling & 5 & 4 \\
		& hc\textsubscript{rnd} & \underline{\textbf{3}} & 1 \\
		& hc\textsubscript{sal+rnd} & 2 & 1 \\
		& ga\textsubscript{rnd} & \underline{\textbf{6}} & 3 \\
		& ga\textsubscript{sal+rnd} & 5 & 3 \\
		
		\bottomrule
		
	\end{tabular}
	\label{table:discussion:ablation-driving}
\end{table*}

When generating roads for the driving environment, we make sure that they are challenging enough for the agent under test. Indeed, we constrain the generation to roads that have at least three curves, one of which must be with a rotation angle of at least $120\degree$. However, the driving agent under test might experience failures also when roads do not obey such constraints. To test this hypothesis, we executed  \tool  with random seeds instead of failure seeds (as  failure seeds obey the constraints by construction) and the sampling approach. \autoref{table:discussion:ablation-driving} shows the number of failures found when executing the failure search with constraints  enabled (i.e., column \textit{w/ Constr.}) and  disabled (i.e., column \textit{w/o Constr.}). Results show that for sampling, hc\textsubscript{sal+rnd} and ga\textsubscript{sal+rnd} there is no statistical significant difference between the number of failures generated by the respective techniques with and without road constraints. However, hc\textsubscript{rnd} and ga\textsubscript{rnd} find significantly more failures when constraints are enabled, showing that restricting the inputs during the search can be beneficial to discover more failures.

\subsection{Regressor-based vs Classifier-based Surrogate Model}

Due to the low number of failures in Parking and Driving in the  setting for RQ\textsubscript{4}, we resorted to a regressor-based surrogate model to guide the search towards failure configurations. We observe that in Parking, the regressor-based surrogate model is as effective as the classifier-based surrogate model. On the other hand, the former is significantly better than the latter in both Humanoid (in terms of higher number of failures, better input diversity and output coverage) and Driving (in terms of higher number of failures and better input diversity).

One explanation of this phenomenon is in the function we use to compute the continuous values for each episode in the different case studies. In Parking, we use the episode length as guidance towards failures, which is not more effective than using a binary classifier. Indeed, the episode length only gives an indirect guidance towards a failure; a long episode might point to configurations where the vehicle is simply far from the parking target spot. On the other hand, in Humanoid, we track the height of the robot throughout the episode; the lower the minimum height the higher the chance that the initial configuration is challenging for the DRL agent under test. Likewise, in Driving we use the minimum cross track error in an episode, which guides the regressor-based surrogate model towards configurations with challenging curves.

In conclusion, the regressor-based surrogate model outperforms the classifier-based one in two out of three case studies, while being equivalent to the classifier-based surrogate model in Parking. In such contexts (see the RQ\textsubscript{4} macro-column in \autoref{table:empirical-evaluation:case-studies}), the number of failures is low (Parking and Driving) or training failures are ineffective (Humanoid), and a continuous-valued function provides a finer-grained signal when training the surrogate model w.r.t. boolean values. However, it requires domain knowledge of the task at hand to define a continuous-valued function that measures how challenging a certain environment configuration is for the agent. %

\subsection{Threats to Validity}

\subsubsection{Internal Validity} A threat to internal validity may come from an unfair comparison of the considered approaches. We gave the same search budget to all approaches, and we generated the same number of environment configurations. Moreover, we considered the same DRL agents for each approach and executed the tests on the same environments.

\subsubsection{External Validity} Using a limited number of subjects poses an external validity threat, in terms of generalizability of our results. To mitigate such threat, we chose three environments which are widely used in the DRL community and have different characteristics that challenge the capabilities of each failure search approach.

\subsubsection{Conclusion Validity} A conclusion validity threat may come from the wrong interpretation of the results due to random variations and inappropriate use of statistical tests. We mitigated this threat by executing each failure search approach multiple times, as well as repeating multiple times the execution of environment configurations on non-deterministic environments. When computing diversity, we executed clustering multiple times to account for the randomness of the $k$-means algorithm. Moreover, we compared the different approaches, both in terms of number of failures and in terms of their diversity, using rigorous statistical tests such as the Mann-Whitney U Test for computing the $p$-value and the Vargha-Delaney metric to measure the effect size. 

\subsubsection{Reproducibility} In terms of reproducibility, we publish our replication package~\cite{replication-package}, making our evaluation repeatable and our results reproducible.

\section{Conclusion and Future Work} \label{sec:conclusion}

Our approach to test DRL agents uses the interaction data produced by the DRL agents during training to train a surrogate model --- i.e., a classifier --- on failure and non-failure (i.e., pass) environment configurations. Then, it uses the failure prediction output of the surrogate model, as a fitness function to be maximized, to achieve high failure-exposure capabilities of the generated environment configurations while saving computation time. Our empirical results show that our search-based approach is able to generate 50\% more failures than the state-of-the-art sampling approach and that such failures are more diverse in all case studies (on average, 78\% more diverse regarding input diversity, and 74\% more diverse regarding output diversity). When training failures are effective at testing time, our results show that \tool generates failures that are complementary. On the other hand, when training failures are ineffective at testing time, \tool triggers significantly more, and more diverse failures. In our future work we plan to increase the diversity of the generated failures by incorporating it into the fitness function.

	\section{Acknowledgements}
This work was partially supported by the H2020 project PRECRIME, funded under the ERC Advanced Grant 2017 Program (ERC Grant Agreement n. 787703).
	
	\balance
	\bibliographystyle{acm}
	\bibliography{paper}
	
	\clearpage
	\appendix
	
	\listofsymbols

\end{document}